\documentclass[aoas,preprint]{imsart}

\newcommand{\Var}{\mbox{Var}}

\newcommand{\ind}{\stackrel{\mathrm{ind}}{\sim}}

\usepackage{amsthm}
\usepackage{amsmath}
\usepackage{amssymb, amsfonts}
\usepackage{natbib}
\usepackage{bm}
\usepackage{graphicx}
\usepackage{courier}
\usepackage{color}
\usepackage{floatrow}
\usepackage{placeins}
\usepackage{units}
\usepackage{url}
\usepackage{verbatim}
\usepackage{epstopdf}
\usepackage{booktabs,caption,fixltx2e}
\usepackage[flushleft]{threeparttable}
\usepackage{rotating}
\usepackage{multirow}
\usepackage{listings} 
\usepackage{subcaption}

\arxiv{math.PR/0000000}

\begin{document}

\begin{frontmatter}

\title{Joint Point and Variance Estimation under a Hierarchical Bayesian model for Survey Count Data\protect\thanksref{T1}}
\runtitle{Bayesian Model for Survey Count Data}
\thankstext{T1}{U.S. Bureau of Labor Statistics, 2 Massachusetts Ave. N.E, Washington, D.C. 20212 USA}

\begin{aug}
\author[A]{\fnms{Terrance D.}~\snm{Savitsky}\ead[label=e1]{Savitsky.Terrance@bls.gov}\orcid{0000-0003-1843-3106}},
\author[B]{\fnms{Julie}~\snm{Gershunskaya}\ead[label=e2]{Gershunskaya.Julie@bls.gov}}
\and
\author[B]{\fnms{Mark}~\snm{Crankshaw}\ead[label=e3]{Crankshaw.Mark@bls.gov}}
\address[A]{Office of Survey Methods Research,
U.S. Bureau of Labor Statistics\printead[presep={,\ }]{e1}}

\address[B]{OEUS Statistical Methods Division,
U.S. Bureau of Labor Statistics\printead[presep={,\ }]{e2,e3}}
\end{aug}

\begin{abstract}
We propose a novel Bayesian framework for the joint modeling of survey point and variance estimates for count data. The approach incorporates an induced prior distribution on the modeled true variance that sets it equal to the generating variance of the point estimate, a key property more readily achieved for continuous data response type models.  Our count data model formulation allows the input of domains at multiple resolutions (e.g., states, regions, nation) and simultaneously benchmarks modeled estimates at higher resolutions (e.g., states) to those at lower resolutions (e.g., regions) in a fashion that borrows more strength to sharpen our domain estimates at higher resolutions.  We conduct a simulation study that generates a population of units within domains to produce ground truth statistics to compare to direct and modeled estimates performed on samples taken from the population where we show improved reductions in error across domains. The model is applied to the job openings variable and other data items published in the Job Openings and Labor Turnover Survey administered by the U.S. Bureau of Labor Statistics.
\end{abstract}

\begin{keyword}
\kwd{Bayesian hierarchical models}
\kwd{Small Area Estimation}
\kwd{Count data}
\kwd{Stan}
\end{keyword}

\end{frontmatter}

\section{Introduction} \label{motivation}

Count data response variables are commonly measured by government surveys; for example, the American Community Survey administered by the U.S. Census Bureau counts the population below a poverty threshold for household domains indexed by geography (e.g., census tracts).  The U.S. Census Bureau administer the Consumer Expenditures surveys of consumer units (independent households) for the U.S. Bureau of Labor Statistics (BLS) that include count variables related to local and regional locations of the consumer units.  BLS administers surveys and a census instrument of business establishments related to total employment and its components (e.g., job openings, hires, separations).  

As with surveys conducted for continuous data response types, surveys that include count data responses aggregate respondent-level counts, such as total employment for a business establishment respondent, to a collection of domains (such as state-by-industry classification) and produce both a point estimate and an estimated variance statistic for each domain. Small domain estimation models for the continuous response type that jointly model the point estimates and the estimated variances for the domains exist within both frequentist and Bayesian frameworks; see, for example, \citet{maiti2014a} and \citet{sugasawa2017a}. These models borrow strength from the underlying correlations among the domain estimates to provide de-noised model-based estimators that are characterized by lower mean squared errors. The inferential goal for these models is to extract model-smoothed point and variance estimates for publication to their data users. It is important to note that the domain-indexed variances are not known, but estimated, such that small domain models provide an opportunity to enhance the quality of both point and variance estimates through their joint estimation since they are typically correlated.

Bayesian models for continuous data point and variance estimates are easily designed such that the mean of the marginal likelihood for the estimated variances represents a denoised “true” variance.  The true variance, in turn, is set to be the “generating” variance for the noisy point estimate in its likelihood centered around the estimated true mean value \citep{sugasawa2017a}; for example, suppose $v_{d}$ represents the estimated sampling variance for domain $d \in (1,\ldots,N)$ associated with \emph{continuous} response, $y_{d}$.  In the case of unknown, latent true domain variance, $\sigma_{d}^{2}$, one may impose a likelihood, $v_{d} \mid \sigma_{d}^{2} \ind f(\sigma_{d}^{2})$ with mean $\sigma_{d}^{2}$. One typically chooses the conditional likelihood, $y_{d} \mid \theta_{d},\sigma_{d}^{2} \ind \mathcal{N}(\theta_{d},\sigma_{d}^{2})$, where $\mathcal{N}(\cdot)$ denotes the normal distribution. We see that the variance in the conditional likelihood for continuous $y_{d}$ is readily and naturally set to equal the latent true variance, $\sigma_{d}^{2}$. This connection between the point and variance estimates where the true modeled variance is set as the generating variance of the noisy point estimate ties together the likelihoods for the point and variance estimates in a single model framework. 

We are not aware of a (small area) model for count data in the small estimation literature where the estimated sampling variance is modeled jointly with the direct point estimate such that the generating variance of the direct point estimate is set equal to the mean of estimated sampling variance (where we interpret the mean as the latent ``true" domain variance). In their recent comprehensive review article of small estimation methods, \citet{Sugasawa2020} note the possibility to model the point estimate with a non-normal distribution and more broadly discuss the use of generalized linear models. They do not, however, explicate a count data model that incorporates estimated variances.  Similarly, \citet{rao2015small} discuss over-dispersed Poisson models for count data, but none that incorporate estimated domain variances.  \citet{Tzavidis2015-sx} develop a Poisson small area model for count data, but assume the domain variance is equal to the mean of the Poisson likelihood for the domain point estimate. So, they do not input domain variances, at all.  

The literature does, however, provide a recent example where \citet{bradley2016a} construct a joint model for geographically-indexed point and variance estimates under a count data response. They define a Poisson likelihood such that the model conditional variance (of the point estimate) is defined as $\Var\left(y_{d}\mid x_{d}\right) = \exp\left(\lambda_{d}\right)$ for count data response, $y_{d}$, associated to domain $d \in (1,\ldots,N)$; where $x_{d}^{{}}$ is a set of covariates and $\lambda_{d}$ is the log-mean parameter.  By contrast, under a normal likelihood with mean $\lambda_{d}$ and variance $\varphi_{d}$ for logarithm, $\log(v_{d})$, of true variance $v_{d}$ of $y_{d}$, the associated mean is $\mathbb{E}\left(v_{d}\mid x_{d}\right) = \sigma^{2}_{d} = \exp\left(\lambda_{d} + \sigma^{2}_{d}/2\right) \neq \Var\left(y_{d}\mid x_{d}\right)$.  Although \citet{bradley2016a} utilize a random effect by specifying a likelihood for the log-variance, this construction does not produce a true variance that is equal to the generating variance of the count data response. All to say, the literature is more limited for small area models for count data that incorporate estimated variances and there are no implementations to our knowledge that ensure the $\sigma_{d}^{2} = \Var\left(y_{d}\mid x_{d}\right)$.  Perhaps the reason for the limited literature focused on count data models for small area estimation is that for many datasets domain level counts are sufficiently large to approximate with a continuous data distribution, though we have mentioned some data examples above with count data variables that express low counts for some domains.

This paper, by contrast, constructs a joint model for a count data point estimate $y_{d}$ and its estimated variance $v_{d}$ where the modeled true variance $\sigma_{d}^{2}$ (the mean of the conditional likelihood for $v_{d}$) is set equal to the variance $\Var(y_{d}\mid x_{d})$ of the point estimate likelihood. We extend a multiplicative random effect in our model specification for the point estimate as suggested by \citet{zhou2012a} for non-survey data where there is no associated variance estimate. They discuss that a Poisson-Lognormal prior set-up better fits the data with more appealing large sample theoretical properties as compared to the Negative Binomial model  because the more flexible formulation of the former allows the data to learn a higher degree of over-dispersion. We extend their Poisson-Lognormal set-up in our Bayesian hierarchical model framework to indirectly induce a prior on the true variance of the likelihood for the estimated variance such that the true variance equals the generating variance for the point estimate.

Our formulation further leverages a Bayesian hierarchical probability model construction by including the variable of interest at multiple resolutions (e.g., nested geographic levels, such as states, regions, nation). The model simultaneously benchmarks modeled point estimates at higher resolutions (e.g., states) to those at lower resolutions (e.g., regions) in a fashion that sharpens the estimation quality at higher resolutions.  Traditional benchmarking discussed in \citet{rao2015small}, by contrast, is often performed as a second step after modeling is completed such that it tends to add back some of the variance removed by modeling. We avoid this loss of efficiency by including the benchmarking as part of performing estimation at multiple resolutions in a single step as is done in \citet{savitsky2016}.

\subsection{Job Opening and Labor Turnover Survey (JOLTS) Motivating Dataset}
This paper was motivated by the JOLTS survey conducted by the U.S. Bureau of Labor Statistics. JOLTS measures dynamic trends in the labor market by tracking job openings, hires and separations, among other variables, on a monthly basis.  The survey is conducted nationally with the intent to provide a national-level estimator for a collection of industries (defined based on the North American industry classification codes (NAICS)). Users, however, desire to have state-level estimates of these labor market dynamic variables for each industry. A major challenge to produce state-level estimates of JOLTS variables is the small number of surveyed business establishments in many states; in fact, in some industries there are states that may not have any sample responses in a given month. It is cost prohibitive for BLS to increase their sample size and to use blocking by state in order to support state level estimation.   Our goal in this paper is to model the collection of state-by-industry point estimates and variances constructed from the national survey to extract more efficient, higher quality estimators and to impute estimated values for state-level domains with no underlying sample. The Bayesian hierarchical model that we construct in the sequel for each industry will simultaneously impute missing point estimates and variances for those states excluded from the sample in any given month.

The remainder of this paper is organized, as follows:  Section \ref{model} provides the mathematical formulation of our joint model for state-level point and variance estimates for each month in each industry.We then extend this cross-sectional by-month model to a time-series construction that jointly models a collection of state-indexed time-series for each of the point and variance estimates in each industry. We design a simulation study in Section \ref{sec:simulation} that generates respondent level population of count data and constructs true values for point estimates over a collection of domains. Our simulation design then takes a sample of respondents and produces domain-based point and both true and estimated variances of direct point estimates for the sampled respondents in each domain. We compare the MSE performances of the sample-based direct estimator and our modeled estimator based on the population ground truth. In Section \ref{sec:application}, we proceed to apply our model to provide JOLTS state-based point and variance estimates and we illustrate the smoothing property of our models. We conclude with a brief discussion in Section \ref{sec:discussion}. 

\section{Model for Count Data Point and Variance Estimates}\label{model}
\subsection{Cross-sectional Model Under Assumed Known Variances}\label{sec:modvarknown}
We proceed to describe the formulation of a cross-sectional model for the joint estimation of count data point in a given month.  We utilize the structure of JOLTS data to describe our model, for ease-of-understanding.  Domains in JOLTS are defined by intersections of states and industries. We consider separate models by industry, each estimated over the collection of states. For each domain $d \in (1,...,N)$ we observe sample based estimates, $y_{d}^{{}},$ and respective estimates of their variances, $v_{d},$ where $N$ denotes the total number of domains in a given industry.  

We construct a model for a count data response, rather than treating point estimate, $y_{d}$, as continuous because the JOLTS variables of interest, such as job openings, are often characterized by very small counts for a given industry and state such that the conditional likelihood is expected to be very skewed (unlike a symmetric normal distribution). Our model will specify a Poisson-lognormal model that allows for an over-dispersed marginal likelihood for point estimate $y_{d}$ for domain $d \in (1,\ldots,N)$ where the data may estimate the variance, $\Var(y_{d}\mid x_{d}) \geq \mathbb{E}(y_{d}\mid x_{d})$.  We assume that the constructed domain variances, $(v_{d})_{d=1}^{N}$, are known such that we treat them as fixed.  So, we specify a likelihood for $y_{d}\mid x_{d}$ and set its variance equal to $v_{d}$, which we see below is not trivial.

\subsubsection{Poisson-lognormal Mixture Likelihood Formulation}

We allow for overdispersion through a Poisson-lognormal joint likelihood with: 
\begin{equation} \label{ylike}
y_{d}^{{}}|\theta _{d}^{{}},{{\varepsilon }_{d}}\ind\mbox{Poisson}\left( \theta _{d}^{{}}{{\varepsilon }_{d}} \right),						
\end{equation}
where the use of latent random effects $\varepsilon_{d}$ with a mean fixed to $1$ allows for the variance to be greater than or equal to the mean by specifying the following distribution for the latent likelihood,
\begin{equation} \label{elike}
\varepsilon _{d}^{{}}|\varphi _{d}^{{}}\ind\mbox{LN}\left( -0.5\varphi _{d}^{2},\varphi _{d}^{2} \right).
\end{equation}
Together, Equations~\ref{ylike} and \ref{elike} for observed direct sample-based estimates $y_{d}^{{}}$ (of job openings, hires, separations, or other JOLTS items) follow a lognormal scale mixture of Poisson distributions parameterized so that mean $E\left( y_{d}^{{}}|\theta _{d}^{{}} \right)=\theta _{d}^{{}},$ where $\theta _{d}^{{}}$ represents the parameter of interest. We accomplish setting the mean of the mixture likelihood to $\theta_{d}$ through our use of $-0.5\varphi_{d}^{2}$ in Equation~\ref{elike} that restricts the prior mean of $\varepsilon_{d}$ to be $1$.  If we had instead chosen a Gamma distribution for $\varepsilon_{d}$ the resulting likelihood after marginalizing over $\varepsilon_{d}$ would have produced a closed-form negative binomial distribution which is \emph{not} the case for our Poisson-lognormal mixture.  We, nevertheless, select the lognormal instead of the Gamma because \citet{zhou2012a} highlight that the lognormal has proven more flexible for the modeling of heavy tails that we express in the JOLTS data to the presence of domains with very small counts.  

\subsubsection{Linking Model for Conditional Mean, $\theta_{d}$}

The conditional mean parameter, $\theta _{d}$, are used to borrow strength across domains and is constructed with the formulation  
\begin{equation} \label{eq:offset}
\theta _{d}^{{}}={{X}_{d}}\exp \left( {{\lambda }_{d}} \right),
\end{equation}
where ${{X}_{d}}$ is an ``offset”. The use of an offset allows specification of the regression model for a normalized rate, $\exp \left( {{\lambda }_{d}} \right)$ which allows the data to estimate correlations among domains for smoothing among domains of different sizes.  The use of a magnitude offset is typical for count data models; for example, in estimating disease prevalance \citep{bda3}. The offset ${{X}_{d}}$ is assumed known and does not contain error. In application to JOLTS, we use the employment level, derived from the Current Employment Statistics (CES) survey conducted by the Bureau of Labor Statistics, as the offset. The total employment values are typically much larger in magnitude that the JOLTS variables (e.g., job openings, quits, hires) and the rate of JOLTS variables to total employment composes a natural ratio in a similar fashion to the number with a disease over the total population in disease mapping \citet{bda3}.    Although the CES-based ${{X}_{d}}$ is an estimate, it is based on a much larger sample than the JOLTS estimate, and so we ignore the variance associated with estimation of${{X}_{d}}$.

We model ${{\lambda }_{d}}$ in Equation~\ref{eq:offset} on the log-scale such that $\lambda_{d} \in \mathbb{R}$ and we may specify a normal distribution prior distribution. We specify a linking model for $\lambda_{d}$ in Equation~\ref{eq:link} that allows ``borrowing strength” across domains from a set of covariates with,
\begin{equation}\label{eq:link}
{{\lambda }_{d}}\sim{\ }\mathcal{N}\left( \beta x_{d}^{{}},{{\tau }^{2}} \right).
\end{equation}
The prior specifies that rate ${{\lambda }_{d}}$ follows a normal distribution, centered at $\beta x_{d}^{{}}$ with variance ${{\tau }^{2}};$ $x_{d}^{{}}$ is a set of covariates and $\beta $ is a vector of regression coefficients. Hyperparameters $\beta $ and ${{\tau }^{2}}$ are ``global” in the sense that their values are shared for all domains.

\subsubsection{Setting Conditional Variance of $y_{d}$ to Equal Known Variance, $v_{d}$}
The Poisson-lognormal mixture likelihood of Equations~\ref{ylike} and \ref{elike} produces the marginal variance,
\begin{equation}
\Var\left(y_{d}\mid \theta_{d},\varphi_{d}\right) = \theta_{d} + \theta_{d}^{2}\left(\exp(\varphi_{d}^{2}) - 1\right),
\end{equation}
where the marginal variance is a function of parameters $(\theta_{d},\varphi_{d})$.  By construction, $v_{d} = \Var(y_{d}\mid X_{d})$, so we need to set the marginal variance (after integrating out $\varepsilon_{d}$) of the Poisson-lognormal mixture equal to $v_{d}$ (treated as known and fixed) with,
\begin{equation}\label{eq:vequal}
    v_{d} = \Var\left(y_{d}\mid \theta_{d},\varphi_{d}\right) = \theta_{d} + \theta_{d}^{2}\left(\exp(\varphi_{d}^{2}) - 1\right).
\end{equation}

Our inferential interest is in $\theta_{d}$, the conditional mean for $y_{d}$ and we specify the linking model for $\theta_{d}$ to borrow strength among domains to produce a smoothed estimator.  So, we accomplish setting $v_{d}$ to be the marginal variance for $y_{d}$ by solving for $\varphi_{d}^{2}$ in Equation~\ref{eq:phiequal} to achieve,
\begin{equation}\label{eq:phiequal}
    \varphi_{d}^{2} = \log \left(\frac{v_{d} - \theta_{d}}{\theta_{d}^{2}} + 1\right),
\end{equation}
where we have \emph{induced} a prior on $\varphi_{d}$ through our linking model distribution imposed on $\lambda_{d}$ (where we recall Equation~\ref{eq:offset}).  In other words, unlike the typical set-up in Bayesian models, we do not directly set a prior distribution for $\varphi_{d}$ but induce it through $\theta_{d}$ and the functional relationship of Equation~\ref{eq:phiequal} to guarantee that Equation~\ref{eq:vequal} is achieved.  As discussed in the introduction, \citet{Sugasawa2020} mention the possibility for use of non-normal likelihoods and generalized linear models, but do not include the joint modeling of point estimates and variances for count data.  Similarly is the case for \citet{rao2015small}.  All to say, ours is the first treatment of a count data model for domain level data that enforces the variance condition of Equation~\ref{eq:vequal}.

\subsection{Model Extension for Joint Modeling of Point Estimates and Variances, $(y_{d},v_{d})$}\label{sec:modvarunknown}

\subsubsection{Likelihood for Observed Variances, $v_{d}$}
We next address the case where the true, underlying domain variances are unknown such that we construct an additional likelihood statement for the observed variances, $v_{d}$, centered on the true, latent variances.  

We denote the true latent variance of sample-based point estimate  $y_{d}^{{}}$ by $\sigma _{d}^{2}$ such that $\sigma _{d}^{2}= \Var\left( y_{d}^{{}}|\theta _{d}^{{}},{{\varphi }_{d}} \right)$.   We achieve this equality by altering Equation~\ref{eq:vequal} to,
\begin{equation}\label{eq:sigequal}
    \sigma_{d}^{2} = \Var\left(y_{d}\mid \theta_{d},\varphi_{d}\right) = \theta_{d} + \theta_{d}^{2}\left(\exp(\varphi_{d}^{2}) - 1\right).
\end{equation}

Note, however, that true sampling variances $\sigma _{d}^{2}$ are not observed. Instead, we observe estimated variances $v_{d}^{{}}$. We choose to work with observed squared coefficient of variation, $cv_{d}^{2}={v_{d}^{{}}}/{y_{d}^{2}}\;$, as doing so allows us to better identify both ${{\varphi }_{d}}$ and $\theta _{d}^{{}}$ by avoiding a multiplicative formulation for the induced prior on $\sigma _{d}^{2}$.  We select a gamma distribution prior for $cv_{d}^{2}$ with mean 
\begin{equation}\label{eq:r_d}
r_{d}^{2}=\frac{\sigma _{d}^{2}}{\theta _{d}^{2}}=\frac{1}{\theta _{d}^{{}}}+\left( \exp \left( \varphi _{d}^{2} \right)-1 \right).
\end{equation}
We specify the prior precision of  $cv_{d}^{2}$ to depend on sample size $n_{d}^{{}}$ and on a random scale hyperparameter ${{a}_{0}}$ that represents a latent concentration property of the population that affects the prior precision of $cv_{d}^{2}$:
\begin{equation}\label{eq:cv_d}
cv_{d}^{2}|{{a}_{0}},r_{d}^{2}\ind\mbox{Gamma}\left( 0.5{{a}_{0}}{{n}_{d}},{0.5{{a}_{0}}{{n}_{d}}}\frac{1}{r_{d}^{2}}\; \right).
\end{equation}
Under this Gamma distribution likelihood, the observed $cv_{d}^{2}$ concentrates to a greater degree on the truth, $r_d^2$, if the domain sample size is relatively large. 

\subsubsection{Prior distribution for Overdispersion, $\varphi_{d}^{2}$}
The latent random effects ${{\varphi }_{d}}$ of Equation~\ref{elike} are interpreted as overdispersion of the Poisson with mean $\theta_{d}.$  We suppose that overdispersion is driven mostly because observed $y_{d}$'s are sample-based estimates (rather than population measurements). Hence, it is natural to assume that the value of parameter ${{\varphi }_{d}}$ depends on the (domain) sample size ${{n}_{d}}$. We let
\begin{equation}\label{eq:phi}
{{\varphi }_{d}}\sim\mathcal{N}_{+}\left( {{\gamma }_{0}}\frac{1}{\sqrt{{{n}_{d}}}},0.1 \right),
\end{equation}
where $\mathcal{N}_{+}(\cdot)$ denotes a half normal distribution for some hyperparameter ${{\gamma }_{0}}$, where this prior specifies a smaller overdispersion for domains with relatively larger sample sizes because relatively larger values of $n_{d}$ result in smaller values of $\varphi_{d}$ which, in turn, produce a lower variance for $\epsilon_{d}$.

Note that we induce a prior on the latent true variance, $\sigma _{d}^{2}$, from Equation~\ref{eq:sigequal} through our direct priors on ${{\varphi }_{d}}$ and ${{\lambda }_{d}}$ specified in Equations~\ref{eq:link} and \ref{eq:phi}, respectively, to ensure that $\sigma _{d}^{2}= \Var\left( y_{d}^{{}}|\theta _{d}^{{}},{{\varphi }_{d}} \right)$; that is, we do not directly specify a prior distribution for $\sigma_{d}^{2}$.

\subsection{Model Extension to Incorporate Regional Data}
We have now fully specified the portion of our hierarchical model for the state-level. We simultaneously model likelihoods for point and variance estimates at the regional level (where for a given industry each region nests a collection of contiguous states) since these aggregated survey estimates are more reliable. We proceed to constrain the modeled point estimates at the state level to sum to those at the region level, which accomplishes simultaneous benchmarking of the states to regions.

In addition, sample based point estimates $y_{r}^{{}}$ and respective estimated variances  $v_{r}$ are also available for regions, $r=1,...,R.$  The United States is subdivided into four regions that nest the states, so $R=4.$ 

To the degree that the regional estimates can be viewed as more reliable compared to the domain estimates, this part of the model serves as a denoised “benchmarking constraint” for domain-level estimates. Yet, because the benchmarking is performed simultaneously with estimation, it produces a lower variance estimate than separately estimating the domain modeled estimates followed by a subsequent benchmarking step. The benchmarking of modeled state estimates to modeled regional estimates also provides a practical benefit by adding stability to estimation of the model parameters. The model for region $r$ is specified as:
\begin{description}
\item[Mixture likelihood for $y_{r}$]
    \begin{subequations}
    \begin{align}
    y_{r}^{{}}|\theta _{r}^{{}} &\ind \mbox{Poisson}\left( \theta _{r}^{{}}\varepsilon _{r}^{{}} \right),~\theta _{r}^{{}}=\sum\limits_{d\in r}{\theta _{d}^{{}}}\\
    \varepsilon _{r}^{{}}|\varphi _{r}^{{}} &\sim{\ }\mbox{LN}\left( -0.5\varphi _{r}^{2},\varphi _{r}^{2} \right)
    \end{align}
    \end{subequations}
\item[Benchmarking step for $\theta_{r}$]
    \begin{equation}\label{eq:bench}
    \theta_{r} = \mathop{\sum}_{d \in r}\theta_{d}
    \end{equation}
\item[Prior for overdispersion $\varphi_{r}$]
    \begin{equation}
    {{\varphi }_{r}}\sim\mathcal{N}_{+}\left( {{\gamma }_{1}}\frac{1}{\sqrt{{{n}_{r}}}},0.1 \right), ~n_{r}^{{}}=\sum\limits_{d\in r}{n_{d}^{{}}}
    \end{equation}
\item[Likelihood for estimated (square of) coefficient of variation $cv_{r}^{2} = v_{r}/y_{r}^{2}$]
    \begin{equation}
    cv_{r}^{2}|{{a}_{1}},r_{r}^{2}\ind\mbox{Gamma}\left( 0.5{{a}_{1}}{{n}_{r}},{0.5{{a}_{1}}{{n}_{r}}}\frac{1}{r_{r}^{2}}\; \right),
    \end{equation}
\end{description}
where $r_{r}^{2}=\frac{1}{\theta _{r}^{{}}}+\left( \exp \left( \varphi _{r}^{2} \right)-1 \right).$   We accomplish a benchmarking step in Equation~\ref{eq:bench} where the mean $\theta _{r}^{{}}$ set equal to the sum of the domain-level parameter values over the domains belonging to a given region.  Thus, the model for the domain means, $(\theta_{d})$ are able to borrow strength from the regional $(y_{r})$ through their links to $(\theta_{r})$ in Equation~\ref{eq:bench}.
We emphasize that this benchmarking step is accomplished simultaneously with estimation of model parameters because the benchmarking is performed among parameters rather than the observed, estimated point estimates.  So, the Bayesian implementation of benchmarking further borrows strength and aids in estimation.  By contrast, the benchmarking discussed in \citet{rao2015small} is performed \emph{after} modeling is performed by benchmarking the modeled estimates to external data benchmarks.  This two-step process, by contrast, adds back some of the noise removed in the model estimation, so it reduces estimation quality.

We proceed to specify a further model extension that benchmarks regional estimates, $\theta_{r}$, to national point and variance estimates.  The form of the extension is analogous to that provided for the regions above, so we omit it for brevity and clarity of exposition.

Model hyperparameters are drawn from the following distributions:
$\beta \sim{\ }\mathcal{N}\left( 0,\sigma _{\beta }^{2} \right),$$\sigma _{\beta }^{{}},\tau \sim \mbox{student-t}_{3+}\left( 0,1 \right)$ (half-Student with 3 degrees of freedom), ${{\gamma }_{0}},{{\gamma }_{1}},\sqrt{{{a}_{0}}},\sqrt{{{a}_{1}}},\sqrt{{{a}_{2}}}\sim\mathcal{N}_{+}\left( 0,1 \right).$

\subsection{A Time-series Extension}

In our JOLTS application presented in Section \ref{sec:application}, we compare the cross-sectional model with an extension that models a collection of (domain-indexed) time-series of point and variance estimates with the intent to borrow more strength from an autocorrelation among the monthly estimates.  We use the same model set-up as the cross-sectional Poisson-lognormal mixture, but now index our parameters by \emph{both} domain, $d$ and time index (month), $t$.
\begin{description}
\item[Mixture likelihood for $y_{dt}$]
    \begin{subequations}
    \begin{align}\label{xbegin}
    y_{dt}^{{}}|\theta _{dt}^{{}},{{\varepsilon }_{dt}} &\ind \mbox{Poisson}\left( \theta _{dt}^{{}}\varepsilon _{dt}^{{}} \right)\\
    \varepsilon _{dt}^{{}}|\varphi _{dt}^{{}} &\sim{\ }\mbox{LN}\left( -0.5\varphi _{dt}^{2},\varphi _{dt}^{2} \right)
    \end{align}
    \end{subequations}
\item[Prior distribution for smoothed point estimate $\theta_{dt}$]
    \begin{subequations}
    \begin{align}
    \theta _{dt}^{{}} &={{X}_{dt}}\exp \left( {{\lambda }_{dt}} \right)\\
    {{\lambda }_{dt}} &\sim{\ }N\left( {{\beta }_{t}}x_{dt}^{{}},{{\tau }^{2}} \right)
    \end{align}
    \end{subequations}
\item[Likelihood for estimated (square of) coefficient of variation $cv_{dt}^{2} = v_{dt}/y_{dt}^{2}$]
    \begin{equation}
    cv_{dt}^{2}|{{a}_{2}},r_{dt}^{2}\ind\mbox{Gamma}\left( 0.5{{a}_{2}}{{n}_{dt}},{0.5{{a}_{2}}{{n}_{dt}}}/{r_{dt}^{2}}\; \right),
    \end{equation}
\item[Prior for overdispersion $\varphi_{r}$]
    \begin{equation}\label{eq:priorphit}
    {{\varphi }_{dt}}\sim{\ }\mathcal{N}_{+}\left( {{\gamma }_{t}}{{x}_{\varphi ,t}},\sigma _{\varphi }^{2} \right)
    \end{equation}
\item[Nonparametric autoregressive priors of order $1$ for $(\beta_{t},\gamma_{t})$]
    \begin{subequations}\label{eq:arprior}
    \begin{align}
    \beta_{t} &\sim \mathcal{N}\left(\beta_{t-1},\sigma_{\beta}^{2}\right)\\
    \gamma_{t} &\sim \mathcal{N}\left(\gamma_{t-1},\sigma_{\gamma}^{2}\right).
    \end{align}
    \end{subequations}
\end{description}
We construct ${x}_{\varphi ,t}=\left( {{\mathbf{1}}_{N}},\frac{1}{\sqrt{{{\mathbf{n}}_{t}}}} \right),~{{\mathbf{n}}_{t}}={{\left( {{n}_{1,t}},...,{{n}_{N,t}} \right)}^{T}}$ in Equation~\ref{eq:priorphit} similarly to the cross-sectional model such the degree of overdispersion is inversely proportional to $\sqrt{n_{dt}}$.  The prior distributions of Equation~\ref{eq:arprior} is formulated as a dynamic linear model of order $1$ \citep{west2013a} with normally distributed innovations that represents a non-parametric local smoother that allows the data to estimate varying patterns of autocorrelation.

Analogous models are formulated for regions and for the total. 

\section{Simulation Study}\label{sec:simulation}

Our goal in this section is to evaluate the model performance when we know the truth (for both a population indexed by units and domains in which the units nest) under a procedure that draws samples from a known population to produce domain estimates for an observed sample that we then compare to the population ground truth. 

We construct a simulation scenario that resembles important characteristics of real data and the survey sampling mechanism. We first generate a finite population stratified by ``geography” and a ``size class” (e.g., discretized categories for number of employees for a business establishment unit) indicator and construct domain true statistics.  Secondly, we draw a stratified simple random sample with replacement from the finite population and obtain direct sample-based and model-based estimates for the population domains. We note that all domains are not guaranteed to be sampled as is the case for states in the JOLTS sample where some states have no sample in a given month.  These steps are repeated a large number of times to capture the variation of population generation and the taking of a sample. The estimates are evaluated against domains true finite population totals, where biases and mean squared errors of competing estimators are calculated over the simulations. Please see ~\ref{suppA} in \citet{savitsky:2022} for details about population generation, the taking of a sample and construction of the direct estimator and its associated variance.

We do not generally have information about the true values of the domain variables when applying candidate models to real data in BLS.  Using a fit statistic to assess the performance of models on the real data would be inappropriate because the purpose of the models is to uncover a latent truth, not to reproduce the observed survey estimates.   So, BLS use simulation studies to assess the relative performance of proposed models as compared to direct estimation in realistic situations that mimic the real world setting as a basis for decision making.

We proceed to use our simulated data in a Monte Carlo simulation study to compare the model formulation of Section~\ref{sec:modvarknown} that treats the domain sampling variances as known and fixed with the extension of Section~\ref{sec:modvarunknown} that treats the underlying true domain variances as unknown and specifies a likelihood for the estimated variances.  Both formulations are compared to the direct estimator and we expect to see substantial reductions in estimation error relatively small-sized domains (with a small sample size).

\subsection{Simulation Results}

The models are estimated with the Hamiltonian Monte Carlo version of Metropolis-Hastings posterior sampling of Stan \citep{gelman2015a} where each posterior sampling iteration provides a draw from the joint posterior distribution over the model parameters.  We choose Stan over the use of a Gibbs sampler primarily because of the induced priors for $\varphi_{d}$ and $\sigma_{d}^{2}$ in Equations~\ref{eq:phiequal} and \ref{eq:sigequal}, respectively.  The induced prior distributions are not of closed form (and therefore highly non-conjugate) such that a Gibbs sampler would be relatively inefficient (due to the need to use some form of slice sampling \citep{Neal2003SliceS}).  By contrast, Stan samples the joint parameter space on each posterior sampling (MCMC) iteration in a single sweep that utilizes the unnormalized joint log posterior distribution, so it doesn't care about conjugacy.  Stan tends to also be more efficient than Gibbs sampling in terms of generating a larger effective number of parameter draws per posterior sampling (MCMC) iteration because the partial suppression of random walk Metropolis-Hastings in HMC makes the sampler less sensitive to a posteriori correlation among the parameters.

Each model run employed 5,000 MCMC iterations, with 2,500 iterations reserved for the warm-up and 2,500 iterations are used to compute model estimates as respective posterior means. We present results after $A=200$ Monte Carlo simulations of the finite populations and respective samples. 

We compare the direct estimator that we label as ``direct" with our two model estimators.  We label the cross-sectional model of Section~\ref{sec:modvarunknown} that jointly models the point estimates and variances, $(y_{d},v_{d})$, under an assumption that the true variance is unknown as ``CS".  We also include a simpler model that supposes the estimated variance $v_{d}$ is the known true variance that is treated as fixed, such that we only model the point estimate, $y_{d}$, as specified in Section~\ref{sec:modvarknown}.  We label this model treating the domain variances as known as ``CS-FV".

The usual statistics to evaluate estimators are an empirical bias of estimator $\hat{Y}_{d}^{{}}$, $B_{d}^{{}}\left( \hat{Y}_{d}^{{}} \right)={{A}^{-1}}\sum\limits_{\alpha =1}^{A}{\left( \hat{Y}_{d\left( \alpha  \right)}^{{}}-Y_{d\left( \alpha  \right)}^{{}} \right)},$ and the square root of the mean of squared errors (RMSE),  $RMSE_{d}^{{}}\left( \hat{Y}_{d}^{{}} \right)=\sqrt{{{A}^{-1}}\sum\limits_{\alpha =1}^{A}{{{\left( \hat{Y}_{d\left( \alpha  \right)}^{{}}-Y_{d\left( \alpha  \right)}^{{}} \right)}^{2}}}},$where $\hat{Y}_{d}^{{}}$’s are respective estimators, direct sample-based or model-based; $Y_{d\left( \alpha  \right)}^{{}}$ is true finite population value and $\hat{Y}_{d\left( \alpha  \right)}^{{}}$ is a realization of  $\hat{Y}_{d}^{{}}$ at simulation run $\alpha $, where $\alpha =1,...,A$.

We present biases and RMSE results in Tables 1-3. Table 1 shows results for domains, grouped by five domain (population size) types, as defined above, where type 5 contains the smallest number of population units and type 1, the largest.  The employment size classes are equally distributed across each domain.  While units (business establishments) containing relatively more employees are over-sampled in each domain, the probability that a domain appears in the sample (through the sampling of at least one nested unit) is proportional to its population size types (number of units).  Domains containing more units are more likely to appear in each (Monte Carlo) sample than domains containing relatively few units.  The number of units sampled within each domain is random such that it varies across realized samples.

\begin{table}[]
\centering
\caption{Bias and RMSE results for domains, grouped by five domain (population size) types (based on $A=200$ runs) where ``Direct" denotes the direct estimator, ``CS-FV", the cross-sectional model under known true variances and ``CS", the cross-sectional model under unknown true variances.}
\label{tab:table1}
\begin{tabular}{p{1cm} p{1cm} p{1cm}   r r r r r r r r}
\multirow{2}{1cm}{domain type} & 
\multirow{2}{1cm}{units per sample} & 
\multirow{2}{1cm}{ave no of samples} & 
& & &
& & & & \\
 &&& \multicolumn{3}{c}{Bias}  &  \multicolumn{3}{c}{RMSE} & \multicolumn{2}{c}{ratio of RMSEs} \\
 &  &  &  \multirow{2}{*}{Direct} & \multirow{2}{*}{CS-FV} & \multirow{2}{*}{CS}  & 
 \multirow{2}{*}{Direct} & \multirow{2}{*}{CS-FV} & \multirow{2}{*}{CS} &  \multirow{2}{*}{CS-FV} & \multirow{2}{*}{CS}    \\
 &&&&&&&&&&\\
 \hline
1 & 31.7 & 200.0 & 1,494 & 983 & -39 & 83,126 & 67,157 & 62,370 & 0.81 & 0.75 \\
2 & 24.6 & 200.0 & -365 & -1,096 & -604 & 64,401 & 47,732 & 43,006 & 0.74 & 0.67 \\
3 & 16.4 & 200.0 & -617 & -1,058 & -973 & 49,855 & 39,159 & 30,334 & 0.79 & 0.61 \\
4 & 8.1 & 199.8 & -440 & 369 & 95 & 35,176 & 29,636 & 16,679 & 0.84 & 0.47 \\
5 & 1.5 & 111.7 & 9 & 1,390 & 129 & 6,445 & 5,385 & 2,014 & 0.84 & 0.31 \\
Overall &  & 182.3 & 17 & -6 & -318 & 56,990 & 44,967 & 39,022 & 0.79 & 0.68
\end{tabular}
\end{table}

\begin{table}[]
\centering
\caption{Bias and RMSE results, regional (based on $A=200$ runs) where ``Direct" denotes the direct estimator, "CS-FV", the cross-sectional model under known true variances and "CS", the cross-sectional model under unknown true variances.}
\label{tab:table2}
\begin{tabular}{ p{1.5cm}   r r r r r r r r}
\multirow{2}{1cm}{units per sample} & 
& & &
& & & & \\
 & \multicolumn{3}{c}{Bias}  &  \multicolumn{3}{c}{RMSE} & \multicolumn{2}{c}{ratio of RMSEs} \\
   &  \multirow{2}{*}{Direct} & \multirow{2}{*}{CS-FV} & \multirow{2}{*}{CS}  & 
 \multirow{2}{*}{Direct} & \multirow{2}{*}{CS-FV} & \multirow{2}{*}{CS} &  \multirow{2}{*}{CS-FV} & \multirow{2}{*}{CS}    \\
 &&&&&&&&\\
 \hline
163 & 11,838 & 9,011 & 9,381 & 173,344 & 138,994 & 121,371 & 0.80 & 0.70  \\
163 & 4,301 & 7,326 & -304 & 175,199 & 147,740 & 128,409 & 0.84 & 0.73  \\
163 & 6 & -4,864 & -3,085 & 204,108 & 157,705 & 143,692 & 0.77 & 0.70   \\
163 & -14,651 & -11,566 & -20,781 & 256,677 & 213,853 & 196,755 & 0.83 & 0.77   \\
Overall & 373 & -23 & -3,697 & 205,113 & 167,145 & 150,483 & 0.81 & 0.73  \\
\end{tabular}
\end{table}

\begin{table}[]
\centering
\caption{Bias and RMSE results, national (based on $A=200$ runs) where ``Direct" denotes the direct estimator, "CS-FV", the cross-sectional model under known true variances and "CS", the cross-sectional model under unknown true variances.}
\label{tab:table3}
\begin{tabular}{ p{1.5cm}   r r r r r r r r}
\multirow{2}{1cm}{units per sample} & 
& & &
& & & & \\
 & \multicolumn{3}{c}{Bias}  &  \multicolumn{3}{c}{RMSE} & \multicolumn{2}{c}{ratio of RMSEs} \\
   &  \multirow{2}{*}{Direct} & \multirow{2}{*}{CS-FV} & \multirow{2}{*}{CS}  & 
 \multirow{2}{*}{Direct} & \multirow{2}{*}{CS-FV} & \multirow{2}{*}{CS} &  \multirow{2}{*}{CS-FV} & \multirow{2}{*}{CS}    \\
 &&&&&&&&\\
 \hline
817 & 1,494 & -93 & -14,789 & 403,256 & 360,231 & 358,282 & 0.89 & 0.89 \\
\end{tabular}
\end{table}

\begin{figure}
\begin{center}
\includegraphics[scale=0.4]{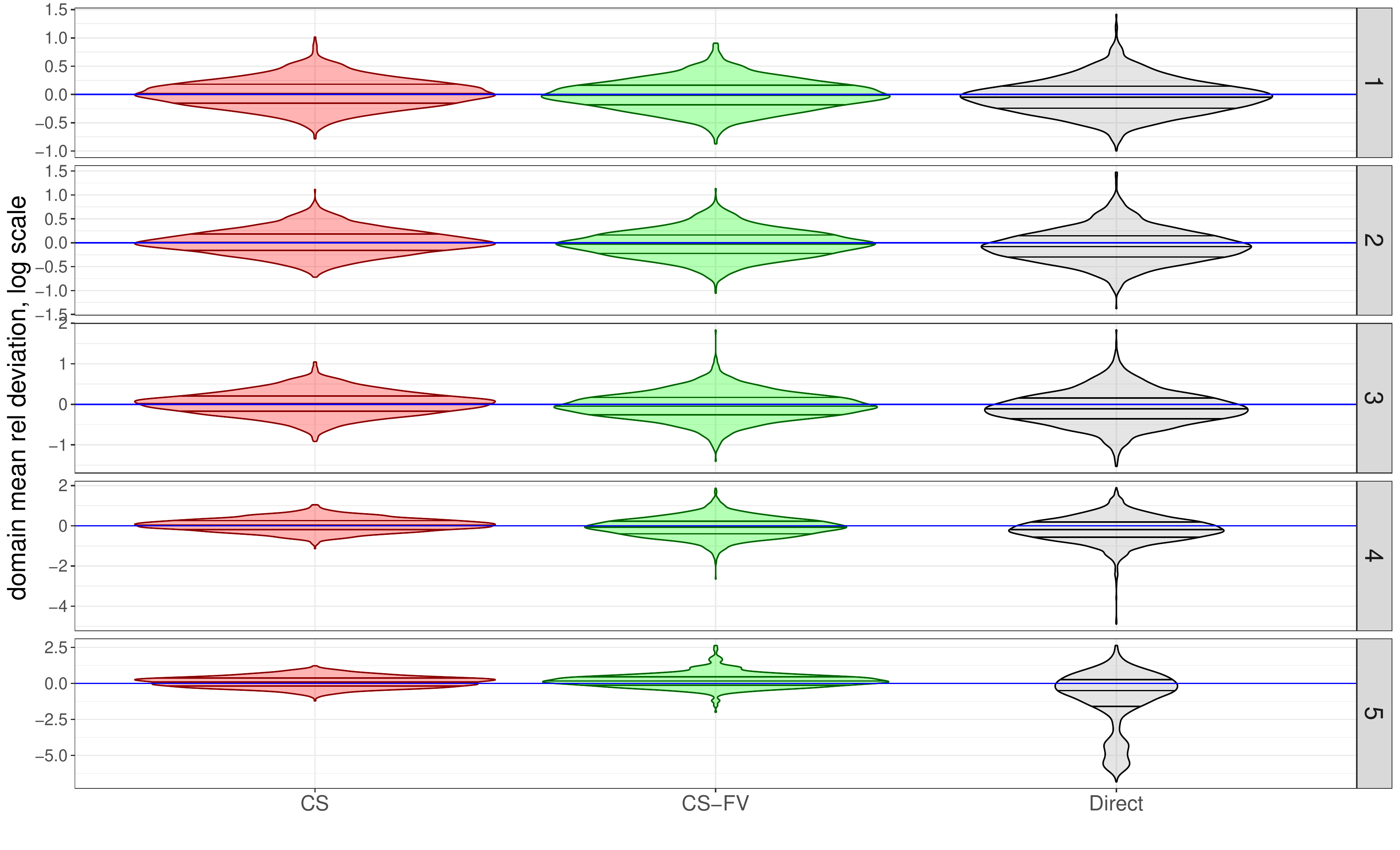}
\caption{Relative errors of point estimates, over simulations, by domain type (where 5 contains the smallest number of population units and type 1 contains the largest), presented on a log scale. ''CS" denotes the cross-sectional model under unknown true variances, ''CS-FV", the cross-sectional model under known true variances and ``Direct" denotes the direct estimator, from left-to-right.}
\label{fig:violinmeanrev}
\end{center}
\end{figure}

\begin{figure}
\begin{center}
\includegraphics[scale=0.4]{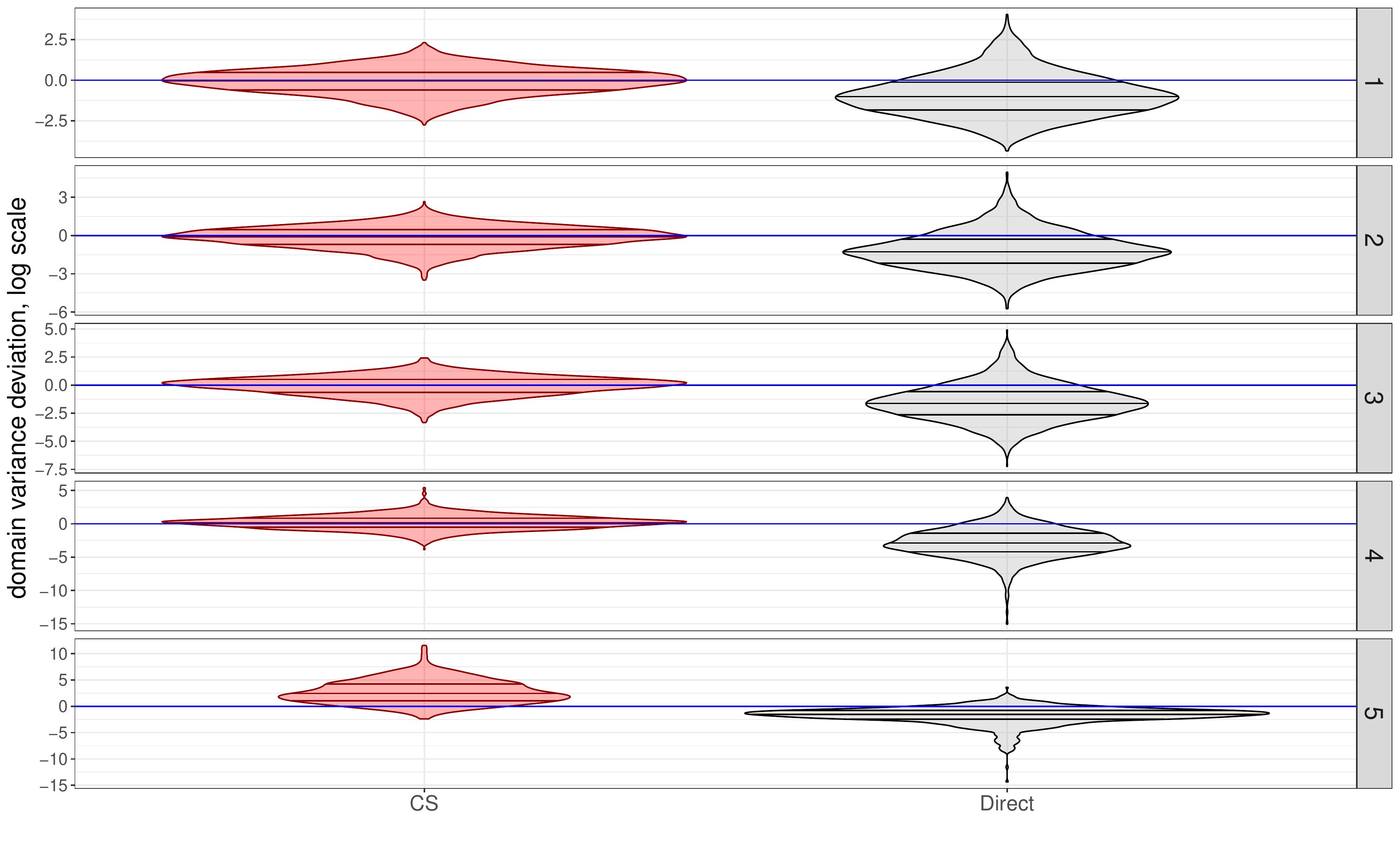}
\caption{Relative errors of variances of direct estimates, over simulations, by domain type (presented on a log scale). ''CS" denotes the cross-sectional model under unknown true variances and ``Direct" denotes the direct estimator, from left-to-right.}
\label{fig:violinvarrev}
\end{center}
\end{figure}

\begin{figure}
     \centering
     \begin{subfigure}[b]{0.45\textwidth}
         \centering
         \includegraphics[width=\textwidth]{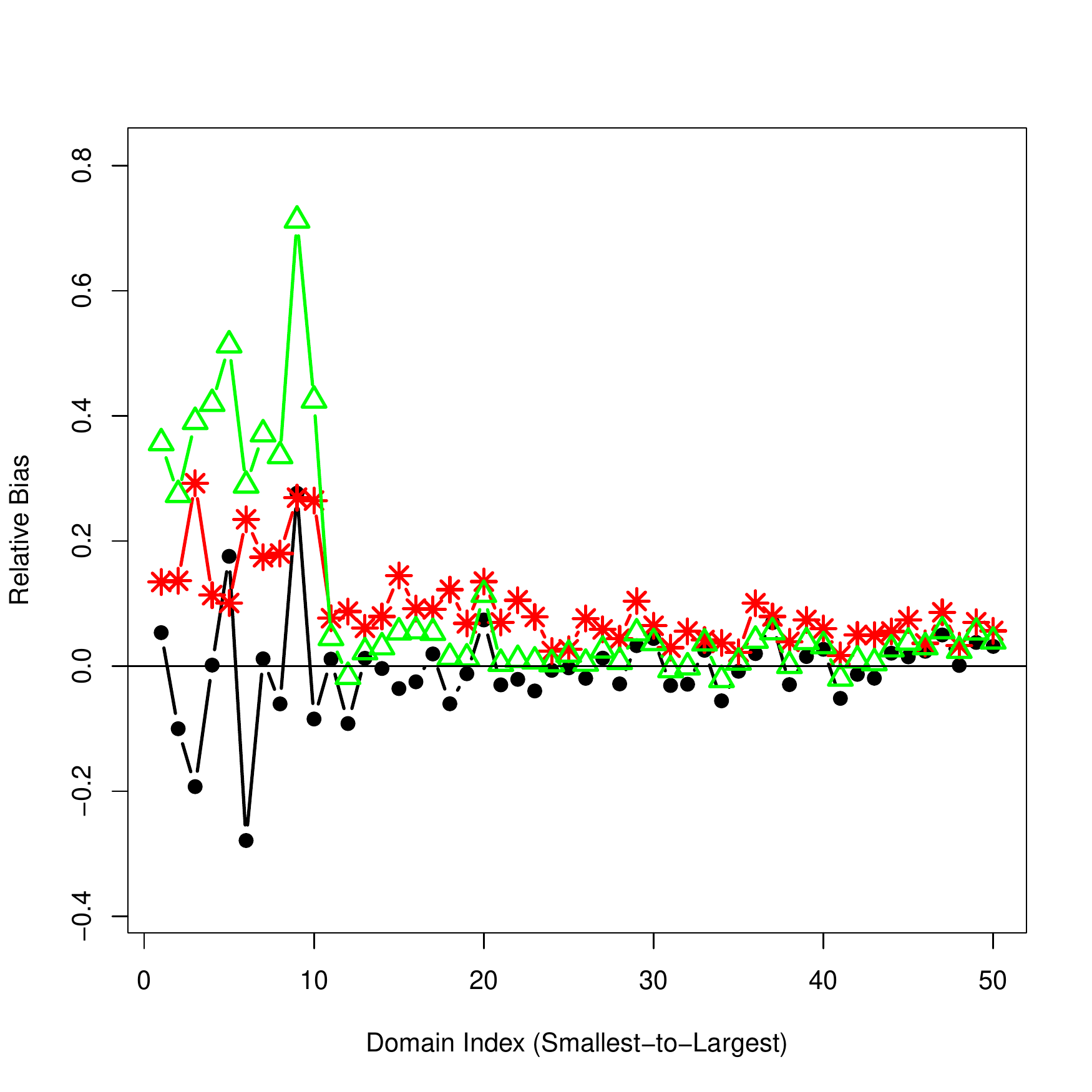}
         \caption{Relative bias}
         \label{fig:relbias}
     \end{subfigure}
     \begin{subfigure}[b]{0.45\textwidth}
         \centering
         \includegraphics[width=\textwidth]{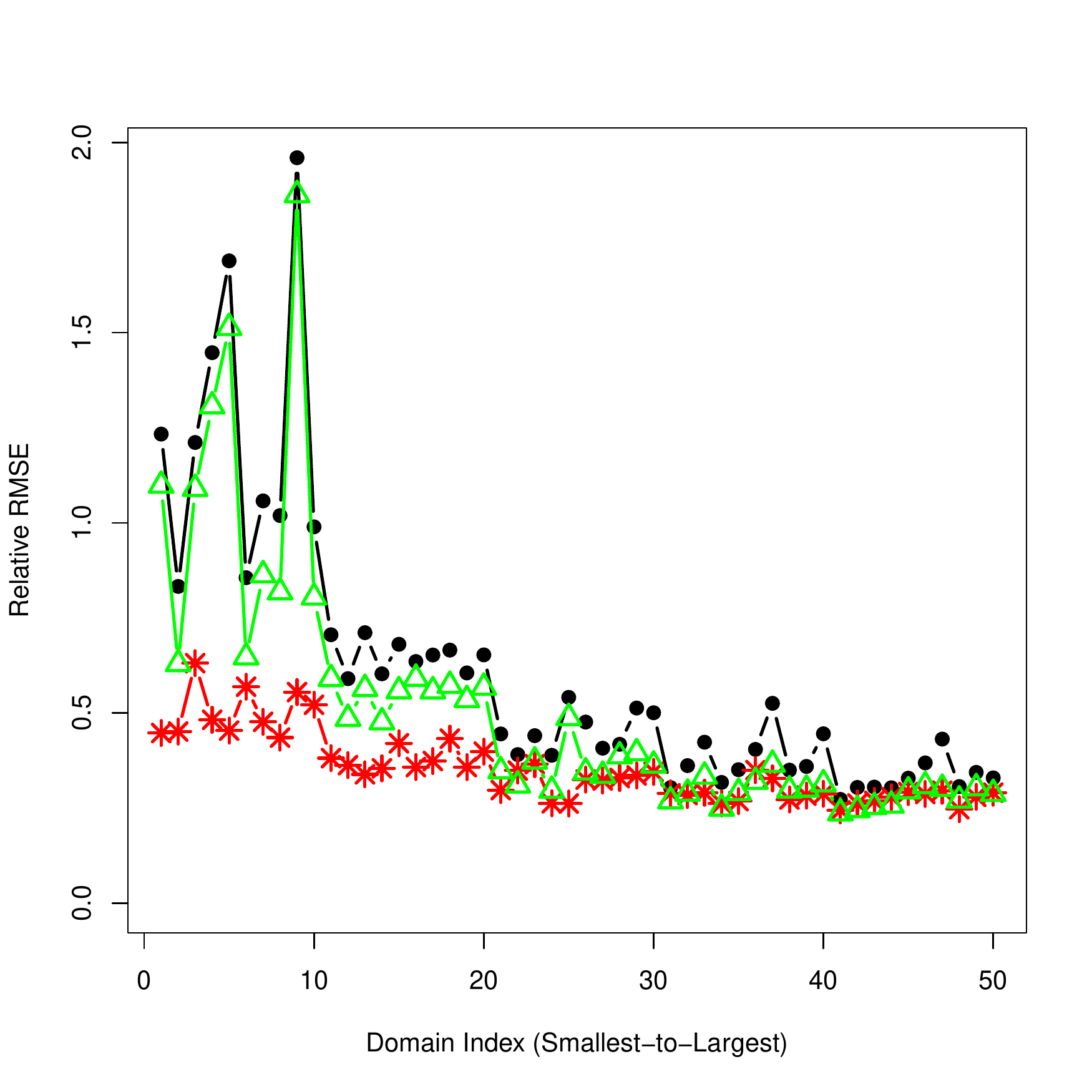}
         \caption{Square root of relative MSE}
         \label{fig:relmse}
     \end{subfigure}
     \caption{Relative biases and square root of relative MSE of point estimates for 50 domains sorted by the number of respondents. The line connecting dots represent our baseline direct estimator, stars represent the CS model, whereas triangles represent the CS-FV model.}
     \label{fig:relerr}
\end{figure}

Table~\ref{tab:table1} presents an average number of sampled domains over the repeated 200 Monte Carlo samples. Note that domains of type 5 are small and are not necessarily represented in each sample. This is reflected in the third column of Table~\ref{tab:table1}, showing an average number of times a domain of type 5 gets into a sample: this value is $111.7$, whereas the larger domains are represented in all $200$ sample realizations.  In the case a domain does not appear in a sample realization in some Monte Carlo iteration, its point estimate is imputed under both the CS and CS-FV models and the sampling variance is also imputed under CS. Table~\ref{tab:table1} reveals that the RMSE ratios between both models and survey direct estimate are always less than 1 for every domain type, but the improvement in error is particularly large for domains of types 4 and 5 at the expense of a relatively small amount of increased bias as compared to the survey-based direct estimator.  The CS model produces a lower RMSE ratio to the direct estimator than does CS-FV for every size class owing to the additional estimation strength borrowed by co-modeling the domain variances. 

Table~\ref{tab:table2} presents results for domains rolled up to “regions”, and Table~\ref{tab:table3} shows results for the overall population total. 

Figure~\ref{fig:violinmeanrev} presents results in a graphical form for the comparison of fit performance between the survey-based direct estimate and the modeled estimate by plotting the distribution over the Monte Carlo iterations of the log of relative deviations constructed as, $\log \left( \frac{\hat{Y}_{d\left( \alpha  \right)}^{{}}}{Y_{d\left( \alpha  \right)}^{{}}} \right)$,  for domains that are present in each sample, over simulations $\alpha $, $\alpha =1,...,A$, where the numerator is the estimated and the denominator is the truth.  A distribution is rendered for each domain type where type 5 represents domains containing the smallest number of population units and type 1, the largest.  Each distribution is presented in the form of a ``violin" plot that mirrors the distribution. Its interpretation is similar to a box plot with additional information for the distribution shape. 
The distribution plots on the right represent the survey-based direct estimates and those in the middle represent the CS-FV model-based posterior means, while those on the left represent the CS model-based posterior means.  The solid horizontal line in each plot panel represents the population true value of the domain point estimates.  We see that while the Monte Carlo distribution over all estimators is centered on the true value, the distributions for the model-based estimators are more concentrated on the true values, with the CS model expressing the smallest variance around the truth.

Figure~\ref{fig:violinvarrev} is an analogous presentation of $\log \left( \frac{\hat{V}_{d\left( \alpha  \right)}^{{}}}{V_{d\left( \alpha  \right)}^{{}}} \right)$ for the sample-based estimates of \emph{variances} of direct estimator and posterior means of the CS model fitted/smoothed estimates of sampling variances ($\sigma_{d}^{2}$).  The solid horizontal line contained in each plot panel of Figure~\ref{fig:violinvarrev} represents the true sampling variance of the point estimator that is approximated over the Monte Carlo iterations by each of the direct sampling variance and the CS model.   In the case that there are no sampled units for small domains (that contain relatively few sampled units) the sampling variances are imputed by the CS model for those missing domains.   The CS-imputed variances represent the quality of the point estimator based on the pattern of borrowing strength from other domains and their underlying estimated variances. The figure reveals that the modeled variances under the CS model (in the left-hand set of plots) are more concentrated than the direct estimates (in the right-hand set of plots), similar to the point estimates.

There is the appearance of a slight positive bias in the fitted sample-based variance estimates (on the left) for domains of types 4 and 5 with fewer population units shown in Figure~\ref{fig:violinvarrev}, but this apparent bias is likely not real.   Our Monte Carlo approximation procedure for the true variance suffers from the exclusion of the null sample realizations for type 4 and 5 domains.   The procedure excludes the null cases for any domain because we are not able to form a sample-based point estimate in any case where the realized sample in any Monte Carlo iteration excludes a domain.  As a result, our Monte Carlo approximation method is expected to under-estimate the true sampling variance for such smaller domains by excluding the null sampling event.  The higher is the probability of the event that a sample realization excludes a domain, the larger would be the under-estimation for our Monte Carlo approximation for the true variance.   Such is likely why we see the appearance of a larger bias for type 5 domains than type 4 (since the probability of a null sample for type 5 domains is higher).

Figure~\ref{fig:relbias} draws a line plot of relative biases $relB_{d}^{{}}\left( \hat{Y}_{d}^{{}} \right)={{A}^{-1}}\sum\limits_{\alpha =1}^{A}{\left( \frac{\hat{Y}_{d\left( \alpha  \right)}^{{}}-Y_{d\left( \alpha  \right)}^{{}}}{Y_{d\left( \alpha  \right)}^{{}}} \right)}$ and Figure~\ref{fig:relmse} draws a line plot of relative root MSE $relRMSE_{d}^{{}}\left( \hat{Y}_{d}^{{}} \right)=\sqrt{{{A}^{-1}}\sum\limits_{\alpha =1}^{A}{{{\left( \frac{\hat{Y}_{d\left( \alpha  \right)}^{{}}-Y_{d\left( \alpha  \right)}^{{}}}{Y_{d\left( \alpha  \right)}^{{}}} \right)}^{2}}}},$ where the x-axis indexes domain labels sorted by the average number of sampled units in each domain from left-to-right.  The line connecting dots represents the direct estimator, the line connecting triangles represents the CS-FV model estimator and the line connecting stars represents the CS model estimator.

Figure~\ref{fig:violinmeanrev} demonstrates that the modeled estimates produce lower true sampling variances than do the survey-based direct estimators.  Figure~\ref{fig:relbias} highlights that the absolute bias is slightly higher, however, particularly for domains with smaller number of sampled units, though the relative MSE improvement for these domains with lower sample sizes are much larger as seen in Figure~\ref{fig:relmse}.

Our model would be expected to outperform direct estimation under any positive correlation between $x_{d}$ and $y_{d}$ because in the worst case setting of no correlation there would be less shrinkage away the direct estimator in the mean parameter, $\theta_{d}$, that we use our model-based estimator.  The correlation between $y_{d}$ and $v_{d}$, however, would provide additional estimation for $\theta_{d}$.  

As the sample size per domain increases, the model estimate will generally contract on the direct estimate, particularly in a flexible model such as ours.   In our experience, one may use the direct estimate in lieu of the model estimate when the number of units per domain is greater than $100$, though some may focus on a target coefficient of variation for favoring the direct estimate.  The direct estimate has the advantage that it is both design unbiased and immune to model misspecification.

In summary, our Poisson-lognormal joint model for $(y_{d},v_{d})$ would be expected to perform well in the case where there are domains with small counts as we see in the JOLTS due to its ability to handle distribution skewness induced by overdispersion.   The performance on domains with larger counts would also be robust, but at the expense of less efficient computation than with a model that treats these domain point estimates as continuous.  We would recommend our model under the set-up where some domains express small counts that induce overdispersion.

\section{Application to JOLTS}\label{sec:application}

We now consider state/industry by-month estimation for the JOLTS job openings variable using a predictor formed from a census instrument that we describe below. Our model estimations for job openings covers the period from January 2014 to December 2018 for up to $52$ states (including Puerto Rico and Washington, DC) depending on the industry being modeled. We compare the smoothing and imputation performances of our 3 model formulations: 1. The cross-sectional (CS) model is separately estimated for each industry and month. In other words, each model is fitted on a set of States in a given industry at a given month; 2. The cross-sectional model under fixed and known variances (CS-FV); 3. The multivariate time series (MV) model jointly models the collection of months in the measurement period for the set of states in each industry, unlike the CS model which estimates one month, at-a-time.

Our model formulations, in addition to state levels, includes likelihood contributions for regional and national levels (for each industry). As a result, model-fitted estimates for regions and national estimates for job openings are also available as a by-product of the model. It is expected that at higher levels / lower resolutions model-fitted estimates should be close to respective sample-based estimates.  A benefit of incorporating these lower resolution regional and national sample-based direct estimates into the model is that higher resolution state-level model-fitted estimates are forced to add up to respective model-fitted regional estimates, which in turn add up to the model-fitted national estimates. This construction is similar to traditional benchmarking, except that we do not require additivity exactly to sample-based lower resolution estimates; instead, we require additivity to model-based lower resolution estimates, which, as we have mentioned are close, yet not exactly equal to, sample-based estimates. 

Another feature of the models is that they simultaneously imputes a value for those months where sample-based estimates are not available (i.e., when there is zero number of respondents.) 

\subsection{Synthetic Predictor, $x_{d}$, for JOLTS Estimation}

JOLTS has designed a synthetic estimator for each state-level domain that we use as a predictor in our modeling. We now briefly describe this estimator. The JOLTS synthetic estimator is constructed by leveraging a census instrument of business establishments, the Quarterly Census of Employment and Wages (QCEW), maintained by BLS for the purpose of measuring employment. The QCEW is administered quarterly and processing time results in a lag of many months before results are published. The QCEW does not publish the components of employment measured by JOLTS, but to obtain the synthetic estimator, the necessary JOLTS variables are ``imputed” in a Hot Deck-like procedure to each record in the QCEW (in a 1 year lag to the current month). This is done, for each month, by stratifying the QCEW records by the intersection of employment change trend (e.g., increasing, decreasing, flat), industry classification and employment size. A data record is randomly drawn from the JOLTS survey data within each stratum and the ratio job openings to total employment for that JOLTS data record is used to impute job openings values of every QCEW record (that contains the employment level for some business establishment) in that stratum. The procedure is repeated until the level of job openings is imputed for every record in the QCEW. The sum of such imputed job openings, across the QCEW records, to each state – by - industry (labeled state/industry) domain provides a “synthetic” estimate. These imputed state/industry synthetic estimates constructed for job openings are further benchmarked to the actual JOLTS sample estimates, by region, to ensure the imputed state levels for job openings equates to the actual JOLTS survey openings levels at the regional and national levels. Similar procedure is performed for other JOLTS data items. As noted, such a synthetic estimator is available based on the lagged QCEW data and does not use the current period JOLTS information.

The synthetic predictor encodes historical expectations (on a one year lag) for each state/industry domain, while the “direct” sample-based estimates supply current information. Synthetic estimates are less variable but they may be biased.  By contrast, direct estimates are considered unbiased, but they have large variance when the sample is small. Model-fitted estimates represent a compromise between “synthetic” and “direct” estimates. The model automatically “decides” on the balance between the synthetic and direct components. This “decision” is made based on the relative variability of the direct sample-based domain estimates and quality of the model fit, which, roughly speaking, depends on how well overall synthetic estimates explain sample-based results. 

\subsection{Results}

Figures 4 and 5 illustrate and compare the relative smoothing and imputation performances of our CS, CS-FV and MV model formulations with the sample-based direct estimate.   Each figure shows alternative estimates for the January 2014-December 2018 period for a single state/industry domain. The dotted line represents CS-FV, the dotdash line represents the CS estimates and the dashed line, the MV estimates.  The solid line represents sample-based direct estimates. Figure 4 presents results for a relatively smaller domain with fewer population business establishment units such that we note the monthly gaps in the solid line where there are no sample-based direct estimates available for that state/industry domain from the national sample in those months.   We observe that all of the models, nevertheless, provide imputed values for these missing months.   The models borrow strength among states with similar patterns of job openings, as well as additionally borrowing strength from the associated regional estimates. 
\begin{figure}
\label{fig: ex1_small}
\begin{center}
\includegraphics[scale=0.8]{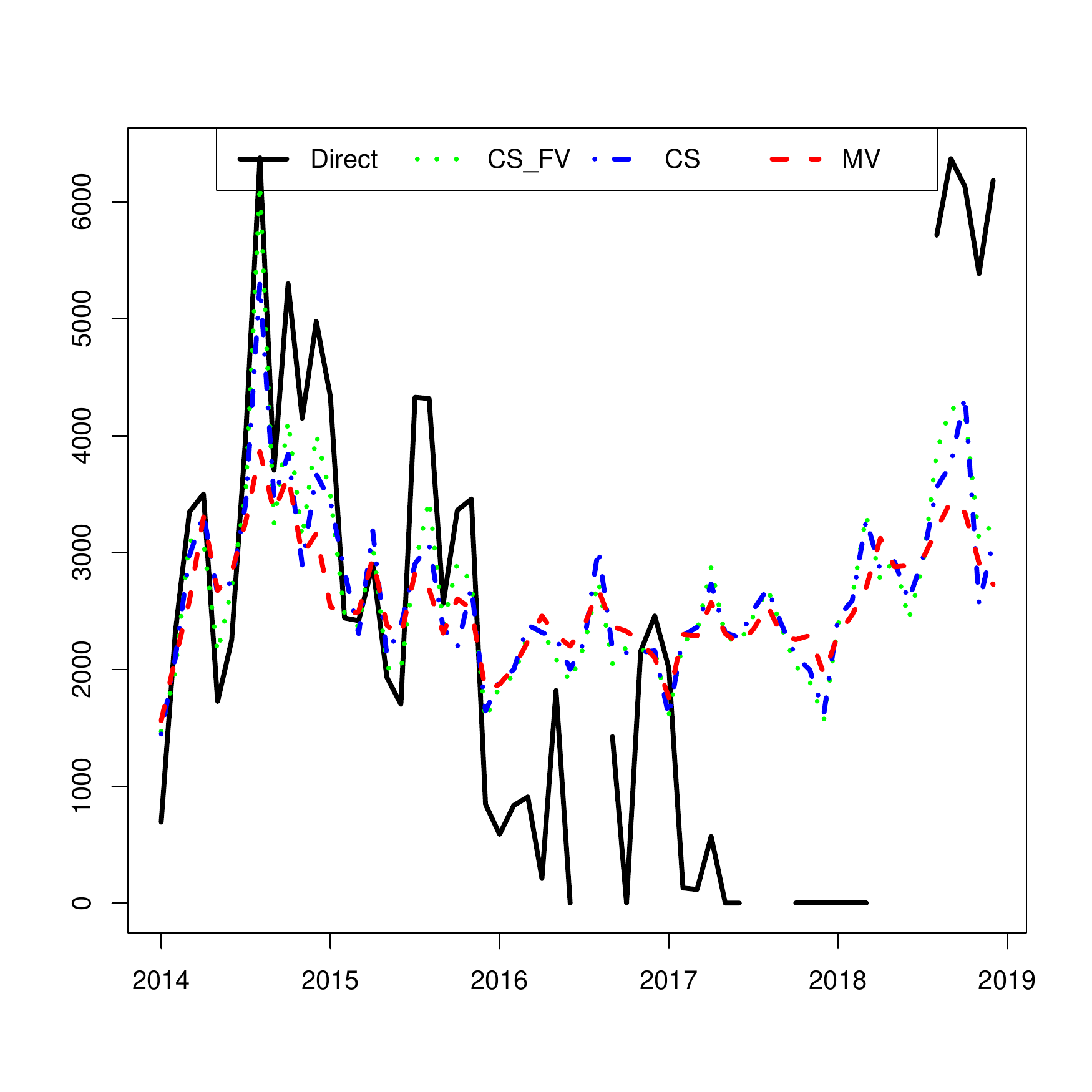}
\caption{Example 1 of domain direct vs cross-sectional (CS) and cross-sectional under known variances (CS-FV) vs multivatiate (MV) time series model fitted estimates, over the period from January 2014 – December 2018, average number of respondents over the period is 4.5}
\end{center}
\end{figure}

\begin{figure}
\label{fig: ex2_large}
\begin{center}
\includegraphics[scale=0.8]{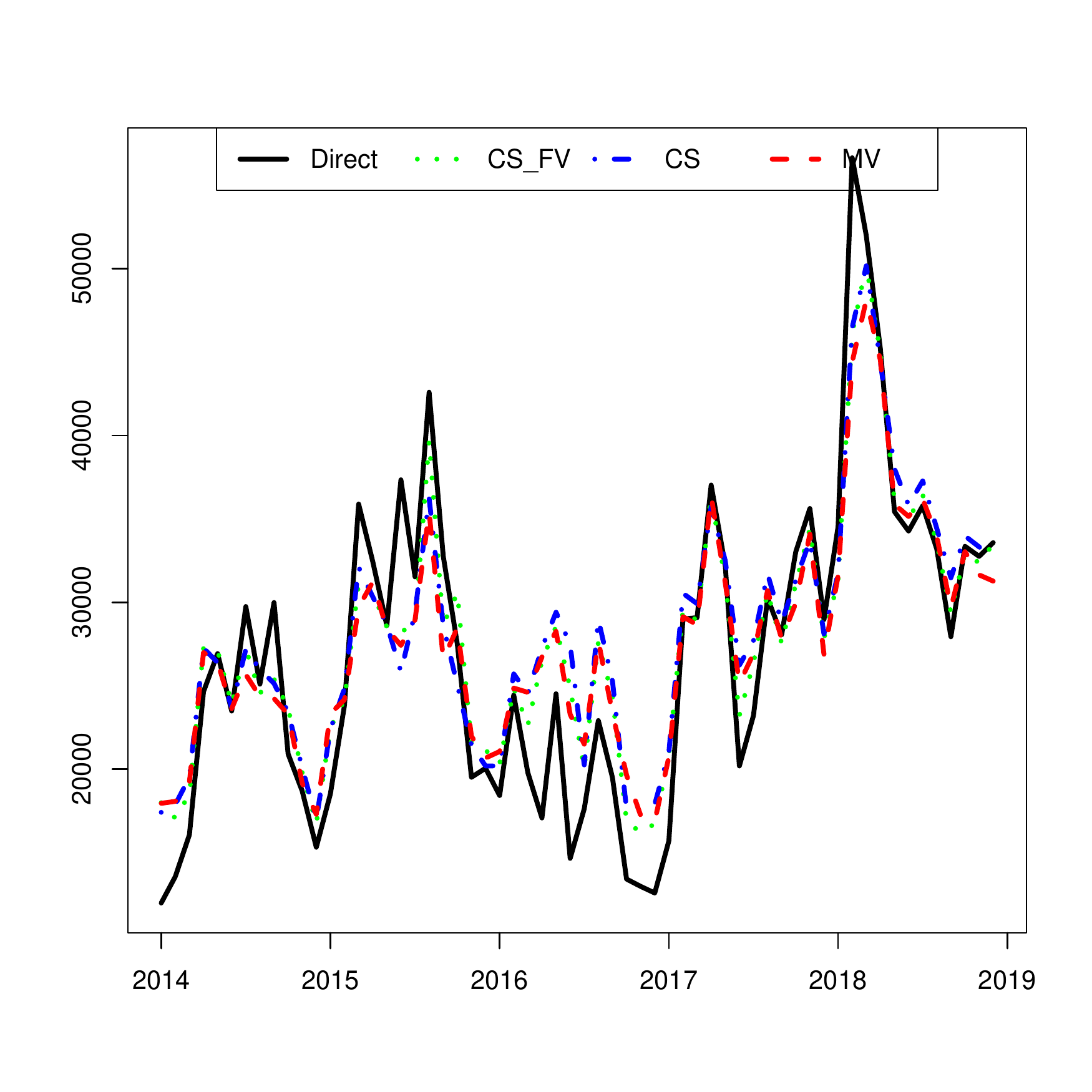}
\caption{Example 2 of domain direct vs cross-sectional (CS) and cross-sectional under known variances (CS-FV) vs multivatiate (MV) time series model fitted estimates, over the period from January 2014 – December 2018, average number of respondents over the period is 53.2}
\end{center}
\end{figure}

The MV borrows further information from the autocorrelation of the monthly estimates co-modeled under this formulation for each state/industry domain.  This additional time-indexed dependence induces a higher degree of over-the-month smoothing.  While the MV model estimation is too computationally-intensive to include within our Monte Carlo simulation study (due to repeated estimations), we ran a few iterations under a simulated dataset generated with autocorrelations and observed that the greater smoothing provided by the MV relative to CS resulted in a slightly, but not notable, improvement in performance as compared to the CS, which is what we also observe in the JOLTS application.  For the models estimated on the mining and logging industry, the CS model estimates in about $500$ seconds, while the $12$ month MV model estimates in about $5500$ seconds.  For this domain containing relatively few business establishment units we see that both models substantially smooth the noisy sample-based direct estimate.

Figure 5 presents the estimation results for a relatively large domain that contains many business establishment population units such that all months are observed.  As expected, the models perform relatively less smoothing of the survey-based direct estimates as these direct estimates are more reliable such that they express lower variances.   For this domain, the CS and MV are relatively coincident with one another with only a small degree of additional smoothing induced by MV.  The CS-FV shows somewhat less smoothing of the direct estimator with more pronounced peaks due to excluding the variances that are useful to strengthen estimation.  The MV model is expected to outperform the CS model in the presence of any autocorrelation so long as the time series pattern doesn't dramatically change due to a ``regime shift" event, such as an economic downturn.  During and after a regime shift event, the correlation structure over time between JOLTS variables would be expected to change such that past month observations before the regime shift would no longer be as useful.  In such a case the CS model would outperform.

\section{Discussion}\label{sec:discussion}

We introduce a novel general Bayesian hierarchical model formulation for the analysis of survey variables of the count data type that extends the automatic property from continuous data model formulations that the estimated true variance is set equal to the generating variance for the point estimate. We set the variance of our Poisson-lognormal construction for the point estimate to the true variance mean of the gamma distribution prior on the observed survey variances. The prior on the true variances is induced by from the prior on the mean and overdispersion parameter of the Poisson-lognormal set-up.

Our modeling approaches further incorporates benchmarking of higher resolution (e.g., state level) estimated totals (which are functions of model parameters) to lower resolution (e.g., regions)  totals simultaneously with their estimation.  This multiresolution construction both sharpens estimates of the higher resolution totals through their nesting in more reliable lower resolution aggregations of domains linked to lower resolution survey-based direct estimates and collapses the usual two-step process of estimating the higher resolution quantities in a model and benchmarking those to lower resolution totals.  Our procedure avoids the reintroduction of noise to lower resolution estimates that typically occurs under the two-step process.

We demonstrated in a simulation study that constructed ground truth estimates for population domains containing varying numbers of underlying units that our count data modeling framework produces lower error estimates than the sample-based direct survey estimates and those improvements are dramatic in domains contain relatively few population units. The model provides these reduction in errors while simultaneously producing imputed estimates for non-sampled domains. Lastly, our application to the JOLTS job openings variable demonstrated that our multivariate model extension provides more smoothing than our base, cross-sectional model by borrowing information from over-the-month correlations.

\begin{supplement}
\sname{Technical Supplement}\label{suppA}
\stitle{Technical Appendices}
\sdescription{The online supplement contains three technical appendices
  with detailed material on the following topics:
\begin{itemize}
  \item[1.] Simulation study detailed design;
  \item[2.] Stan \citep{gelman2015a} scripts for models;
\end{itemize}}
\end{supplement}

\bibliography{JOLTS}
\bibliographystyle{agsm}

\appendix

\section{Details of Population and Sample Design for Simulation Study}
We provide details below for generating a finite population indexed by units, domains and regions.   We further describe how a sample is drawn from the population using a stratified, size-based sample using the average stratum employment as the size variable.  We proceed to describe the ratio estimator we use for our direct survey estimator constructed from the observed sample along with the computation of its survey variance estimator.

\subsection{Configuring the Finite Population: number of units, domains, and regions} 

We set a population of size ${{N}_{P}}=1,000,000$ consisting of $D=50$ domains. There are 5 types of domains, based on their number of population units each domain type contains; namely, ${{N}_{d}}\in \left\{ 39000,30000,20000,10000,~1000 \right\},$$d=1,...,50.$ In this set-up, the 10 largest-sizes domains are constructed with 39,000 population units, whereas the 10 smallest domains are each assigned only 1,000 population units. 

Each domain belongs to one of $R=4$ “regions”: domains $d=1,...,10$ are in region $r=1$; $d=11,...,20.$ are in region $r=2$; $d=21,...,30$ are in region $r=3$; and domains $d=31,...,50$are region $r=4.$ 

\subsection{Constructing Population “Employment Size Classes”}

The population is subdivided into $S=6$ “employment size classes” based on variable $y_{j}^{emp}$ (“employment”). Let size class mean levels be $m_{s}^{emp}=\left\{ 2,10,20,40,100,1000 \right\}$. The total number of population units assigned to each size class is${{N}_{s}}=\left\{ 700000,110000,90000,70000,20000,10000 \right\},$$\sum\limits_{s=1}^{S}{{{N}_{s}}}={{N}_{P}}.$ The “employment” level for each business establishment unit in each size class is generated as a Poisson variable $y_{j}^{emp}\sim{\ }Poisson\left( m_{s}^{emp} \right),j\in s,s=1,...,S.$ Each domain includes approximately equal proportion of units from each size class. The true finite population domain employment level that we generate as $Y_{d}^{emp}=\sum\limits_{j=1}^{{{N}_{d}}}{y_{j}^{emp}}$, is assumed to be known for our modeling.

\subsection{Generating a Sub-employment Variable}

Let predictor, $x_{d}^{{}}\sim{\ }Unif\left( 0.02,0.3 \right),d=1,...,D$ be a domain specific predictor that is assumed to be fixed and known. We set parameters$\sigma _{\lambda }^{2}=0.1$, $\sigma _{\varepsilon }^{2}=1$, $\beta =0.7$ and generate $\lambda _{d}^{{}}\sim{\ }N\left( \beta \log \left( {{x}_{d}} \right),\sigma _{\lambda }^{2} \right),d=1,...,D,$ and  ${{\varepsilon }_{j}}\sim{\ }N\left( -0.5\sigma _{\varepsilon }^{2},\sigma _{\varepsilon }^{2} \right),j=1,...,{{N}_{P}}.$

Population values for a sub-employment variable (e.g., job openings, hires or separations), ${{y}_{j}}$ , are generated from an overdispersed Poisson distribution, as${{y}_{j}}\sim{\ }Poisson\left( m_{j}^{{}} \right),$ with means${{m}_{j}}=y_{j}^{emp}\exp \left( \lambda _{d}^{{}}+\varepsilon _{j}^{{}} \right),j=1,...,{{N}_{P}}$.
True finite population targets are domain totals $Y_{d}^{{}}=\sum\limits_{j=1}^{{{N}_{d}}}{y_{j}^{{}}}$.

\subsection{Sampling Design and Estimation}

We use a stratified simple random sampling with replacement design, where strata are defined by intersections of regions and size classes, $h=1,...,H,H=RS$. Sampling selection probabilities are defined by employment size strata as ${{\pi }_{s}}=\left( 0.00025,\text{ }0.00075,\text{ }0.00125,\text{ }0.0025,0.0075,0.0125 \right)$, which induces respective sampling weights ${{w}_{s}}=\left( 4000,1333,800,400,133,80 \right)$.  Strata that share the same employment size classes across regions are assigned the same sampling selection probabilities. In designing the sampling scheme for this simulation, we strive to produce sampling weights that would resemble those in the actual real data JOLTS application, where the weights range from about 1 to 5000.
The direct sample weighted domain ratio estimator has the form \[\hat{Y}_{d}^{Dir}=Y_{d}^{emp}\hat{R}_{d}^{{}},\] 
where 
\[\hat{R}_{d}^{{}}=\frac{\sum\limits_{j\in {{S}_{d}}}^{{}}{{{w}_{j}}y_{j}^{{}}}}{\sum\limits_{j\in {{S}_{d}}}^{{}}{{{w}_{j}}y_{j}^{emp}}},\]
with $S_{d}^{{}}$ is the set of sampled units in domain $d$ and ${{w}_{j}}$ is a sampling weight of unit $j\in {{S}_{d}}$.

Note that this sampling design does not guarantee that a given domain is represented in the sample. There is a chance that some of the smaller domains would not be included into sample, and thus, their direct estimates are not defined. In this case the model would still provide an estimate for these domains. 

The model input data includes variances of direct estimates, which we compute from the sample using the usual linearization formula of the ratio estimator:  \[\hat{V}_{d}^{Dir}=\sum\limits_{h}^{{}}{\frac{N_{dh}^{2}}{n_{dh}^{{}}}\frac{\sum\limits_{j\in {{S}_{dh}}}^{{}}{{{\left( u_{dj}^{{}}-\bar{u}_{dj}^{{}} \right)}^{2}}}}{n_{dh}^{{}}-1}},u_{dj}^{{}}=y_{j}^{{}}-{{\hat{R}}_{d}}y_{j}^{emp},\bar{u}_{dj}^{{}}=\frac{1}{n_{dh}^{{}}}\sum\limits_{j\in {{S}_{dh}}}^{{}}{u_{dj}^{{}}},\]
where ${{S}_{dh}}$ is a set of sampled units belonging to stratum $h$ and domain $d$.  Associated true sampling variances for each domain are constructed by computing the variance of the point estimate over a collection of Monte Carlo simulations, where each simulation generates a population and takes a sample of units from that population.

The “regional” and “national” level point and variance estimates are derived using analogous formulas. 

\section{Stan Scripts for Selected Models} 
We present Stan scripts for our two major models: 1. Cross-sectional model with unknown true variances; 2. Time series model with unknown true variances.

\subsection{Stan Script for Cross-sectional Model with Unknown True Variance of Section 2.2}

Cross-sectional model:

\begin{lstlisting}
    /* 
      one month at a time
    */
  
  data{
    int<lower=1> N;                     // number of domains
    int<lower=0> y[N];                  // set of N observations (direct sample-based domain estimates)
    row_vector<lower=0>[N] cv2_y;       // observed cv2
    row_vector<lower=0>[N] nResp;       // number of respondents
    int<lower=1> P;                     // number of covariates included in matrix x
    matrix[N,P] x;                      // Design matrix for P predictors 
    row_vector<lower=0>[N] Emp;         // offset (e.g., employment)
    
    int<lower=0> N_obs; 		// Number of non-missing values
    int<lower=0> N_miss; 		// Number of missing values
    int<lower=1> ind_obs[N_obs];     	// Vector of non-missing value indices
    int<lower=1> ind_miss[N_miss];     	// Vector of missing value indices

    int<lower=1> R;                     // number of regions

    matrix[N,R] region;			// region assignment
    int<lower=0> y_r[R];                // set of R observations for regions
    row_vector<lower=0>[R] cv2_y_r;     // observed cv2, regions
    real<lower=0> cv2_y_nat;            // observed cv2, national

  } /* end data block */
  
  
  transformed data{
    vector[P] zros_b;
    matrix[N,P] logx;  
    int<lower=0> y_obs[N_obs];             // non-missing ys
    real<lower=0> nResp_nat;               // number of respondents, national
    int<lower = 0> y_nat;                  // y for national level
    row_vector<lower=0>[R] nResp_r;        // number of respondents, regions
    row_vector<lower=0>[N_obs] cv2_y_obs;  // observed cv2
    row_vector<lower=0>[N_obs] nResp_obs;  // number of respondents, observed

    zros_b       = rep_vector(0,P);
    logx         = log(x);

    y_obs	 = y[ind_obs];
    nResp_obs	 = nResp[ind_obs];
    cv2_y_obs    = cv2_y[ind_obs];

    nResp_r      = nResp * region ;
    y_nat        = sum( y_r );
    nResp_nat    = sum(nResp_r) ;

  } /* end transformed parameters block */
  
  
  parameters{
    real<lower=0> sqrt_shape; 
    real<lower=0> sqrt_shape_r; 
    real<lower=0> sqrt_shape_nat; 

    real<lower=1> vbias; /*** bias in observed variances (underestimates) of the regional estimates **/

    real<lower=0> sigma_lam; 
    row_vector[N] lambda; 
    row_vector[N] log_epsilon_raw;

    row_vector<lower=0>[N] sqrt_phi;
    row_vector<lower=0>[R] sqrt_phi_r;
    real<lower=0> sqrt_phi_nat;
    real<lower=0> phi_beta;


    vector[P] beta;
    vector<lower=0>[P] sigma_b; /* vector of sd parameters for P x P, Lambda */
    
    /* region */ 
    row_vector[R] log_epsilonr_raw;

    /* National */
    real log_epsilon_nat_raw;

    
  } /* end parameters block */
  
  
  transformed parameters{
    vector[N] xb;
    row_vector<lower=0>[N] fitted_y;
    row_vector<lower=0>[N] mean_y;
    row_vector<lower=0>[N_obs] mean_y_obs;
    row_vector<lower=0>[N] fitted_vrnc;
    row_vector<lower=0>[N] fitted_cv2 ;
    row_vector[N] log_epsilon; 
    row_vector<lower=0>[N] epsilon; 
    
    
// phies
    row_vector<lower=0>[N] phi;
    row_vector<lower = 0>[R] phi_r;  //phies for region
    real<lower = 0> phi_nat;  //National

    row_vector<lower=0>[N_obs] fitted_cv2_obs;
    
    /* region */

    row_vector<lower=0>[R] fitted_y_r;
    row_vector<lower = 0>[R] fitted_vrnc_r;
    row_vector<lower=0>[R] fitted_cv2_r;
    row_vector[R] log_epsilonr; 
    row_vector<lower=0>[R] epsilonr; 
    row_vector<lower=0>[R] mean_y_r;

    /* National */
    real<lower=0> fitted_y_nat;
    real<lower = 0> fitted_vrnc_nat;
    real<lower = 0> fitted_cv2_nat;
    real log_epsilon_nat;
    real<lower=0> epsilon_nat;
    real<lower=0> mean_y_nat;


    real<lower=0> shape; 
    real<lower=0> shape_r; 
    real<lower=0> shape_nat; 


    shape       = square(sqrt_shape);
    shape_r     = square(sqrt_shape_r);
    shape_nat   = square(sqrt_shape_nat);

    /* states */
    phi         = square((sqrt_phi));
    phi_r       = square( sqrt_phi_r );
    phi_nat     = square( sqrt_phi_nat );

    log_epsilon = log_epsilon_raw .* (sqrt_phi) - 0.5*phi;
    epsilon     = exp(log_epsilon); 
    xb          = logx * beta ;
    fitted_y    = Emp .* exp(lambda);
    mean_y      = Emp .* exp(lambda ) .* epsilon;
    mean_y_obs  = mean_y[ind_obs];
    fitted_cv2  = inv(fitted_y) + (exp(phi) - 1) ;
    fitted_cv2_obs  = fitted_cv2[ind_obs];
    fitted_vrnc = (fitted_y .* fitted_y) .* fitted_cv2 ;


  
   // add up by regions

    log_epsilonr  = log_epsilonr_raw .* sqrt_phi_r - 0.5*phi_r;
    epsilonr      = exp(log_epsilonr);
    fitted_y_r    = fitted_y * region ;
    mean_y_r      = fitted_y_r .* epsilonr;
    fitted_cv2_r  = inv(fitted_y_r) + (exp(phi_r) - 1) ;
    fitted_vrnc_r = (fitted_y_r .* fitted_y_r) .* fitted_cv2_r ;
    
   // National
    fitted_y_nat    = sum( fitted_y_r );
    log_epsilon_nat = log_epsilon_nat_raw * sqrt_phi_nat - 0.5*phi_nat;
    epsilon_nat     = exp(log_epsilon_nat);


    mean_y_nat      = fitted_y_nat * epsilon_nat;
    fitted_cv2_nat  = inv(fitted_y_nat) + (exp(phi_nat) - 1) ;
    fitted_vrnc_nat = square(fitted_y_nat) * fitted_cv2_nat;

  } /* end transformed parameters block */
  
  model{
    { /* local block for parameters */
      sigma_lam       ~ student_t( 3, 0.0, 1.0 ); 
      lambda          ~ normal( xb, sigma_lam);
      log_epsilon_raw ~ std_normal();
    } /* end local block for parameters */
      
      
    { /* local variable block  */
      sigma_b         ~ student_t( 3, 0.0, 1.0 );
      beta            ~ normal( 0, sigma_b );
    } /* end local variable block */

     sqrt_shape        ~ std_normal();
     sqrt_shape_r      ~ std_normal();
     sqrt_shape_nat    ~ std_normal();

     vbias             ~ normal(0,10);

     //sqrt_phi        ~ std_normal();            
     //sqrt_phi_r      ~ std_normal();             
     //sqrt_phi_nat    ~ std_normal();

     phi_beta 	       ~ std_normal();            
     sqrt_phi 	       ~ normal(phi_beta*inv(sqrt(nResp)),1);            
     sqrt_phi_r        ~ normal(phi_beta*inv(sqrt(nResp_r)),1);            
     sqrt_phi_nat      ~ normal(phi_beta*inv(sqrt(nResp_nat)),1);            
 
     y_obs             ~ poisson(mean_y_obs);
     cv2_y_obs         ~ gamma (0.5*shape*nResp_obs, 
                                0.5*shape*nResp_obs .* inv( fitted_cv2_obs ));
 
     log_epsilonr_raw  ~ std_normal();
     y_r               ~ poisson( mean_y_r);
     cv2_y_r           ~ gamma (0.5*shape_r*nResp_r, 
                                0.5*shape_r*nResp_r .* inv( fitted_cv2_r/vbias ));

     log_epsilon_nat_raw ~ std_normal();
     y_nat               ~ poisson( mean_y_nat);
     cv2_y_nat           ~ gamma (0.5*shape_nat*nResp_nat, 
                                  0.5*shape_nat*nResp_nat * inv( fitted_cv2_nat/vbias ));


  } /* end model{} block */
\end{lstlisting}

\newpage

\subsection{Time-series Model with Unknown True Variances of Section 2.5} 
Time-series model:

\begin{lstlisting}
    /* 
      a collection of months, together
    */
  
  data{
    int<lower=1> N;                     // number of domains
    int<lower=1> T;                     // number of months
    // All N*T length vectors assumed stacked as T, N x 1 columns
    int<lower=0> y[N*T];                // set of T, N x 1 stacked obs 
                                        // (direct sample-based domain estimates)
    row_vector<lower=0>[N*T] cv2_y;     // observed cv2
    row_vector<lower=0>[N*T] nResp;     // number of respondents
    int<lower=1> P;                     // number of covariates included in matrix x
    matrix[N*T,P] x;                    // Design matrix for P predictors 
    row_vector<lower=0>[N*T] Emp;       // offset (e.g., employment)
    
    int<lower=0> N_obs[T]; 		// Number of non-missing values
    // T stacked columns, each of dimension N_obs[t], t = 1,...,T
    // Assume that each set of N_obs[t] contains entries \in (1,..,N) of observed states
    int<lower=1> ind_obs[sum(N_obs)];     	// Vector of non-missing value indices

    int<lower=1> R;                     // number of regions-months

    matrix[N,R] region;			// region assignment 
    int<lower=0> y_r[R*T];              // stack R regions with each set repeated T times
    row_vector<lower=0>[R*T] cv2_y_r;   // observed cv2
    row_vector<lower=0>[T] cv2_y_nat;   // observed cv2

  } /* end data block */
  
  
  transformed data{
    vector[P] zros_b;
    vector[R] ones_R;
    vector[R*T] ones_RT;
    vector[N*T] ones_NT;
    vector[T] ones_T;
    matrix[N,P] logx; 
    int<lower=1> ind_cum_obs[sum(N_obs)];
    int<lower=0> y_obs[sum(N_obs)];            // non-missing ys
    row_vector<lower=0>[T] nResp_nat;          // number of respondents
    int<lower=0> y_r_mat[R,T];                 // render R*T x 1, y_r as a matrix
    int<lower = 0> y_nat[T];
    row_vector<lower=0>[R*T] nResp_r;          // number of respondents
    row_vector<lower=0>[sum(N_obs)] cv2_y_obs; // observed cv2
    row_vector<lower=0>[sum(N_obs)] nResp_obs; // number of respondents
    matrix<lower=0>[N*T,2] x_phi;
    matrix<lower=0>[R*T,2] x_phir;
    matrix<lower=0>[T,2] x_phinat;
    

    zros_b        = rep_vector(0,P);
    ones_R        = rep_vector(1,R);
    ones_T        = rep_vector(1,T);
    ones_RT       = rep_vector(1,R*T);
    ones_NT       = rep_vector(1,N*T);
    
    logx          = log(x);


    // convert indexiing from ind_obs where each set of N_obs[t] indices are \in (1,...N)
    // to positions in a stacked set of vectors with indices in (1,...,(N*T)) 
    // to map to y and cv2, which are stacked
    { /* local block */
      int pos_t     = 1;
      int end_t     = 0;
      int pos_cum_t = 0;
      for(t in 1:T)
      {
        end_t   = pos_t + N_obs[t] - 1;
        for( h in pos_t:end_t)
        {
          ind_cum_obs[h]                    = ind_obs[h] + pos_cum_t; /* \in (1,...,N*T) */
        }
        pos_t     += N_obs[t]; 
        pos_cum_t += N; /* Add N to each N_obs[t] vector of missing state values beginning with month t = 2 */
      }
    } /* end local block */
    
    y_obs	      = y[ind_cum_obs];
    nResp_obs	  = nResp[ind_cum_obs];
    cv2_y_obs   = cv2_y[ind_cum_obs];

    { /* local block */
      int pos_t = 1;
      int posn_t = 1;
      int end_t = 0;
      int endn_t = 0;
      for( t in 1:T )
      {
        end_t                 = t*R;
        endn_t                = t*N;
        nResp_r[pos_t:end_t]  = nResp[posn_t:endn_t] * region;
        y_r_mat[1:R,t]        = y_r[pos_t:end_t];/* stack by column */
        pos_t                 = end_t + 1;
        posn_t                = endn_t + 1;
        y_nat[t]              = sum( y_r_mat[1:R,t] );
      } /* end loop t over T months */
    } /* end local block */

    nResp_nat    = ones_R' * to_matrix(nResp_r,R,T,1);

    x_phi         = append_col(ones_NT,inv(sqrt(nResp))');
    x_phir        = append_col(ones_RT,inv(sqrt(nResp_r))');
    x_phinat      = append_col(ones_T,inv(sqrt(nResp_nat))');

  } /* end transformed parameters block */
  
  
  
  parameters{
    real<lower=0> sqrt_shape; 
    real<lower=0> sqrt_shape_r; 
    real<lower=0> sqrt_shape_nat;
    real<lower=0> sigma_sphi;
    real<lower=0> sigma_sphir;
    real<lower=0> sigma_sphinat;

    real<lower=1> vbias; /*** bias in observed variances (underestimates) of the regional estimates **/

    real<lower=0> sigma_lam; 
    row_vector[N*T] lambda; 
    row_vector[N*T] log_epsilon_raw;

    row_vector<lower=0>[N*T] sqrt_phi;
    row_vector<lower=0>[R*T] sqrt_phi_r;
    row_vector<lower=0>[T] sqrt_phi_nat;
    matrix<lower=0>[T,2] phi_beta_raw;
    vector<lower=0>[2] sigma_phi;
    matrix<lower=0>[T,2] phir_beta_raw;
    vector<lower=0>[2] sigma_phir;
    matrix<lower=0>[T,2] phinat_beta_raw;
    vector<lower=0>[2] sigma_phinat;


    matrix[T,P] beta_raw;
//    cholesky_factor_corr[P] C_b; /* cholesky of correlation matrix for cols of Beta */
    vector<lower=0>[P] sigma_b; /* vector of sd parameters for P x P, Lambda */
    
    /* region */ 
    row_vector[R*T] log_epsilonr_raw;

    /* National */
    row_vector[T] log_epsilon_nat_raw;

    
  } /* end parameters block */
  
  
  transformed parameters{
    matrix[T,P] beta;
    matrix<lower=0>[T,2] phi_beta;
    matrix<lower=0>[T,2] phir_beta;
    matrix<lower=0>[T,2] phinat_beta;
    vector[N*T] xb;
    row_vector<lower=0>[N*T] fitted_y;
    matrix<lower=0>[N,T] fitted_y_mat;
    row_vector<lower=0>[N*T] mean_y;
    row_vector<lower=0>[sum(N_obs)] mean_y_obs;
    row_vector<lower=0>[N*T] fitted_vrnc;
    row_vector<lower=0>[N*T] fitted_cv2 ;
    row_vector[N*T] log_epsilon; 
    row_vector<lower=0>[N*T] epsilon; 
    
    
// phies
    row_vector<lower=0>[N*T] phi;
    row_vector<lower = 0>[R*T] phi_r;  //phies for region
    row_vector<lower = 0>[T] phi_nat;  //National
    row_vector<lower=0>[N*T] fitted_sqrtphi;
    row_vector<lower=0>[R*T] fitted_sqrtphi_r;
    row_vector<lower=0>[T] fitted_sqrtphi_nat;

    row_vector<lower=0>[sum(N_obs)] fitted_cv2_obs;
    
    /* region */

    row_vector<lower=0>[R*T] fitted_y_r;
    row_vector<lower = 0>[R*T] fitted_vrnc_r;
    row_vector<lower=0>[R*T] fitted_cv2_r;
    row_vector[R*T] log_epsilonr; 
    row_vector<lower=0>[R*T] epsilonr; 
    row_vector<lower=0>[R*T] mean_y_r;

    /* National */
    matrix<lower=0>[R,T] fitted_y_r_mat;
    row_vector<lower=0>[T] fitted_y_nat;
    row_vector<lower = 0>[T] fitted_vrnc_nat;
    row_vector<lower = 0>[T] fitted_cv2_nat;
    row_vector[T] log_epsilon_nat;
    row_vector<lower=0>[T] epsilon_nat;
    row_vector<lower=0>[T] mean_y_nat;


    real<lower=0> shape; 
    real<lower=0> shape_r; 
    real<lower=0> shape_nat; 


    shape       = square(sqrt_shape);
    shape_r     = square(sqrt_shape_r);
    shape_nat   = square(sqrt_shape_nat);

    /* states */
    phi         = square((sqrt_phi));
    phi_r       = square( sqrt_phi_r );
    phi_nat     = square( sqrt_phi_nat );
    
    // place an RW(1) prior on gamma, kappa, beta
    // e.g., gamma[i,t] ~ N(gamma[i,t-1],sigma_gam)
    for( p in 1:P )
    {
      beta[,p] = cumulative_sum(beta_raw[,p]) * sigma_b[p];
    } /* end loop p over P regression coefficients in model for lambda */
   
    for( p in 1:2 )
    {
      phi_beta[,p]    = cumulative_sum(phi_beta_raw[,p]) * sigma_phi[p];
      phir_beta[,p]   = cumulative_sum(phir_beta_raw[,p]) * sigma_phir[p];
      phinat_beta[,p] = cumulative_sum(phinat_beta_raw[,p]) * sigma_phinat[p];
    } /* end loop p over intercept and nResp predictors */
    

    log_epsilon = log_epsilon_raw .* (sqrt_phi) - 0.5*phi;
    epsilon     = exp(log_epsilon); 
    { /* local block */
      int beg_t    = 1;
      int beg_tr   = 1;
      int end_t    = 0;
      int end_tr   = 0;
      for( t in 1:T )
      {
        end_t                             = t*N;
        end_tr                            = t*R;
        xb[beg_t:end_t]                   = logx[beg_t:end_t,] * beta[t,]';
        fitted_sqrtphi[beg_t:end_t]       = (x_phi[beg_t:end_t,] * phi_beta[t,]')';
        fitted_sqrtphi_r[beg_tr:end_tr]   = (x_phir[beg_tr:end_tr,] * phir_beta[t,]')';
        fitted_sqrtphi_nat[t]             = dot_product(x_phinat[t,], phinat_beta[t,]) ;
        beg_t                             = end_t + 1;
        beg_tr                            = end_tr + 1;
      }
    }/* end local block */
    
    fitted_y        = Emp .* exp(lambda);
    fitted_y_mat    = to_matrix(fitted_y,N,T,1); /* stack the columns */
    mean_y          = Emp .* exp(lambda ) .* epsilon;
    mean_y_obs      = mean_y[ind_cum_obs];
    fitted_cv2      = inv(fitted_y) + (exp(phi) - 1) ;
    fitted_cv2_obs  = fitted_cv2[ind_cum_obs];
    fitted_vrnc     = (fitted_y .* fitted_y) .* fitted_cv2 ;


  
   // add up by regions

    log_epsilonr    = log_epsilonr_raw .* sqrt_phi_r - 0.5*phi_r;
    epsilonr        = exp(log_epsilonr);
    for( t in 1:T )
    {
      fitted_y_r_mat[1:R,t] = region' * fitted_y_mat[1:N,t];
    }
    fitted_y_r      = to_row_vector(fitted_y_r_mat); /* stack the columns */
    mean_y_r        = fitted_y_r .* epsilonr;
    fitted_cv2_r    = inv(fitted_y_r) + (exp(phi_r) - 1);
    fitted_vrnc_r   = (fitted_y_r .* fitted_y_r) .* fitted_cv2_r ;
    
   // National
    fitted_y_nat        = ones_R' * fitted_y_r_mat;
    log_epsilon_nat     = log_epsilon_nat_raw .* sqrt_phi_nat - 0.5*phi_nat;
    epsilon_nat         = exp(log_epsilon_nat);


    mean_y_nat        = fitted_y_nat .* epsilon_nat;
    fitted_cv2_nat    = inv(fitted_y_nat) + (exp(phi_nat) - 1) ;
    fitted_vrnc_nat   = square(fitted_y_nat) .* fitted_cv2_nat;

  } /* end transformed parameters block */
  
  model{
    { /* local block for parameters */
      sigma_lam       ~ student_t( 3, 0.0, 1.0 ); 
      lambda          ~ normal( xb, sigma_lam );
      log_epsilon_raw ~ std_normal();
    } /* end local block for parameters */
      
      
    { /* local variable block  */
      sigma_b             ~ student_t( 3, 0.0, 1.0 );
      to_vector(beta_raw) ~ std_normal();
    } /* end local variable block */
      

     sigma_phi         ~ student_t( 3, 0.0, 1.0 ); 
     sigma_phir        ~ student_t( 3, 0.0, 1.0 ); 
     sigma_phinat      ~ student_t( 3, 0.0, 1.0 ); 
     sqrt_shape        ~ std_normal();
     sqrt_shape_r      ~ std_normal();
     sqrt_shape_nat    ~ std_normal();
     sigma_sphi        ~ student_t( 3, 0.0, 1.0 ); 
     sigma_sphir       ~ student_t( 3, 0.0, 1.0 ); 
     sigma_sphinat     ~ student_t( 3, 0.0, 1.0 );

     vbias        ~ normal(0,10);

     
     to_vector(phi_beta_raw)      ~ std_normal();
     to_vector(phir_beta_raw)     ~ std_normal();
     to_vector(phinat_beta_raw)   ~ std_normal();

     sqrt_phi 	  	~ normal(fitted_sqrtphi,sigma_sphi);            
     sqrt_phi_r 	  ~ normal(fitted_sqrtphi_r,sigma_sphir);            
     sqrt_phi_nat 	~ normal(fitted_sqrtphi_nat,sigma_sphinat);            
 
     y_obs             ~ poisson(mean_y_obs);
     cv2_y_obs         ~ gamma (0.5*shape*nResp_obs, 
                                0.5*shape*nResp_obs .* inv( fitted_cv2_obs ));
 
     log_epsilonr_raw  ~ std_normal();
     y_r               ~ poisson( mean_y_r );

     cv2_y_r           ~ gamma (0.5*shape_r*nResp_r, 
                                0.5*shape_r*nResp_r .* inv( fitted_cv2_r/vbias ));

     log_epsilon_nat_raw  ~ std_normal();
     y_nat                ~ poisson( mean_y_nat);
     cv2_y_nat            ~ gamma (0.5*shape_nat*nResp_nat, 
                                   0.5*shape_nat*nResp_nat .* inv( fitted_cv2_nat/vbias ));


  } /* end model{} block */

\end{lstlisting}

\end{document}